\documentclass{article}

\usepackage{ifthen}
\usepackage{pictexwd,dcpic}
\usepackage{graphicx}
\usepackage{amsfonts,amsmath,amssymb}

\newcommand{\Fig}[3]
{\bigskip
\begin{figure}\begin{center}
\ifthenelse{\equal{#2}{}}
{\includegraphics[width=200pt,height=200pt,draft]{./pics/#1.pcx}}
{\DMS#2,0{./pics/#1.pcx}}
\caption{{\footnotesize{#3}}}\label{#1}\end{center}
\end{figure}
\bigskip
}

\def\be{\begin{eqnarray}}
\def\ee{\end{eqnarray}}
\def\nn{\nonumber}

\def\p{\partial}

\def\l[{\phantom.[}

\hoffset= - 1.2in

\begin{document}

\hfill IITP/TH-11/15

\bigskip

\centerline{\Large{Colored HOMFLY and Generalized Mandelbrot set
}}

\bigskip

\centerline{{\bf Ya.Kononov$^d$, A.Morozov$^{a,b,c}$}}

\bigskip

{\footnotesize
\centerline{{\it
$^a$ ITEP, Moscow 117218, Russia}}

\centerline{{\it
$^b$ National Research Nuclear University MEPhI, Moscow 115409, Russia
}}

\centerline{{\it
$^c$ Institute for Information Transmission Problems, Moscow 127994, Russia
}}

\centerline{{\it
$^d$ Higher School of Economics, Math Department, Moscow, 117312, Russia
}}
}

\bigskip

\bigskip

\centerline{ABSTRACT}

\bigskip

{\footnotesize

Mandelbrot set is a closure of the set of zeroes of $resultant_x(F_n,F_m)$
for iterated maps
$F_n(x)=f^{\circ n}(x)-x$ in the moduli space of maps $f(x)$.
The wonderful fact is that for a given $n$ all zeroes are not chaotically scattered
around the moduli space, but lie on smooth curves,
with just a few cusps, located at zeroes of $discriminant_x(F_n)$.
We call this phenomenon the {\it Mandelbrot property}.
If approached by the cabling method, symmetrically-colored HOMFLY polynomials
$H^{\cal K}_n(A|q)$ can be considered
as linear forms on the $n$-th  "power" of the knot ${\cal K}$,
and one can wonder if zeroes of $resultant_{q^2}(H_n,H_m)$
can also possess the Mandelbrot property.
We present and discuss such resultant-zeroes patterns in the complex-$A$ plane.
Though $A$ is hardly an adequate parameter to describe the moduli space of knots,
the Mandelbrot-like structure is clearly seen -- in full accord with the vision of
arXiv:hep-th/0501235, that concrete slicing of the Universal Mandelbrot set is not
essential for revealing its structure.


}

\bigskip

\bigskip

\section{Introduction}

The purpose of this paper is to establish, at least phenomenologically,
a clear relation between the two celebrated entities from the "opposite" corners
of theoretical physics: {\bf Mandelbrot sets} \cite{MS}, belonging to the world of
discrete-time dynamical systems, and {\bf colored HOMFLY polynomials} \cite{knotpols},
providing the most important kind of knot invariants.
Formulation and analysis of this relation is made possible by the
algebro-geometric reformulation of Mandelbrot theory in \cite{DMand}
and by the recent progress in understanding of symmetric HOMFLY, originated
in \cite{IMMMfe}.

\bigskip

Mandelbrot set is a famous example of fractal
structure,
provided by consideration of iterated maps $f^{\circ n}(x) = f\Big( f^{\circ(n-1)}(x)\Big)$,
which from the point of view of dynamical systems describe evolution in the discrete time $n$.
If the shape of the map $f(x)$ is characterized by additional parameters (moduli) $c$,
then the distribution of stable orbits
in the $c$-space is described by the famous pattern (generated by \cite{FE} for the basic example
$f_c(x) = x^2+c$):

\bigskip

\centerline{
\includegraphics[scale=0.5]{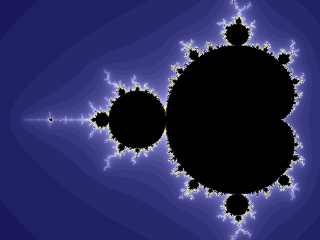}
}
\vspace{-3cm}
\be
\label{Fig1}
\ee
\vspace{1.1cm}

\bigskip

\noindent
From a programmer's perspective, shown in black are the values of complex $c$,
for which the sequence $f_c^{\circ n}(x=0)$ does not tend to $x=\infty$
(the choice of initial point $x=0$ is not always adequate -- see \cite{AndMand} for
improved algorithm).
This, however, explains nothing about peculiar -- and, in fact, universal -- structure
in the picture.
What explains, is its algebro-geometric description in \cite{DMand} (see also \cite{AndMand,DMand2}
and \cite{NLA}).

Namely, consider a shifted iteration $F_n(x) =f^{\circ n}(x)-x$, its zeroes in $x$-space describe
the closed orbits of length $n$.
These orbits are stable, provided $\left|\frac{df^{\circ n}}{dx}\right|<1$ --
and this is true or not, depending on the value of $c$.
The two equations
\be
\left\{\begin{array}{c}
F_n(x)=0\\
\left|\frac{df^{\circ n}}{dx}\right|=1
\end{array}\right.
\label{Mandanalyt}
\ee
describe an almost-everywhere-smooth curve $L_n$ in the complex-$c$ plane
(perhaps, made of several components).
In particular, the central cardioid in the picture is
\be
c = \frac{1}{2}e^{i\theta} - \frac{1}{4}e^{2i\theta}
\ \ \ \ \ \ \Longleftarrow \ \ \ \ \ \
\left\{ \begin{array}{c}
x^2+c-x=0\\
2|x|=1
\end{array} \right.
\ee
The Mandelbrot set (or, if one prefers, its boundary) is actually a collection of these curves.

This is, however, not yet the needed algebro-geometric description.
To get it, we further note that stability of the $n$-th  orbit is lost when a Feigenbaum
cycle-multiplication \cite{Feig} occurs, and what gets stable is the orbit of the order $(kn)$
with some integer $k$.
Such bifurcation happens at zeroes of
$resultant_x(F_n,F_{kn}/F_n)$, which densely populate the curve $L_n$ --
and therefore $L_n$ can be considered as a closure of the set of these zeroes for all $k$.
Cusp singularities of $L_n$ are actually located at zeroes of $discriminant_x(F_n)$.
In this description the central cardioid appears as

\bigskip

\centerline{
\includegraphics[scale=0.5]{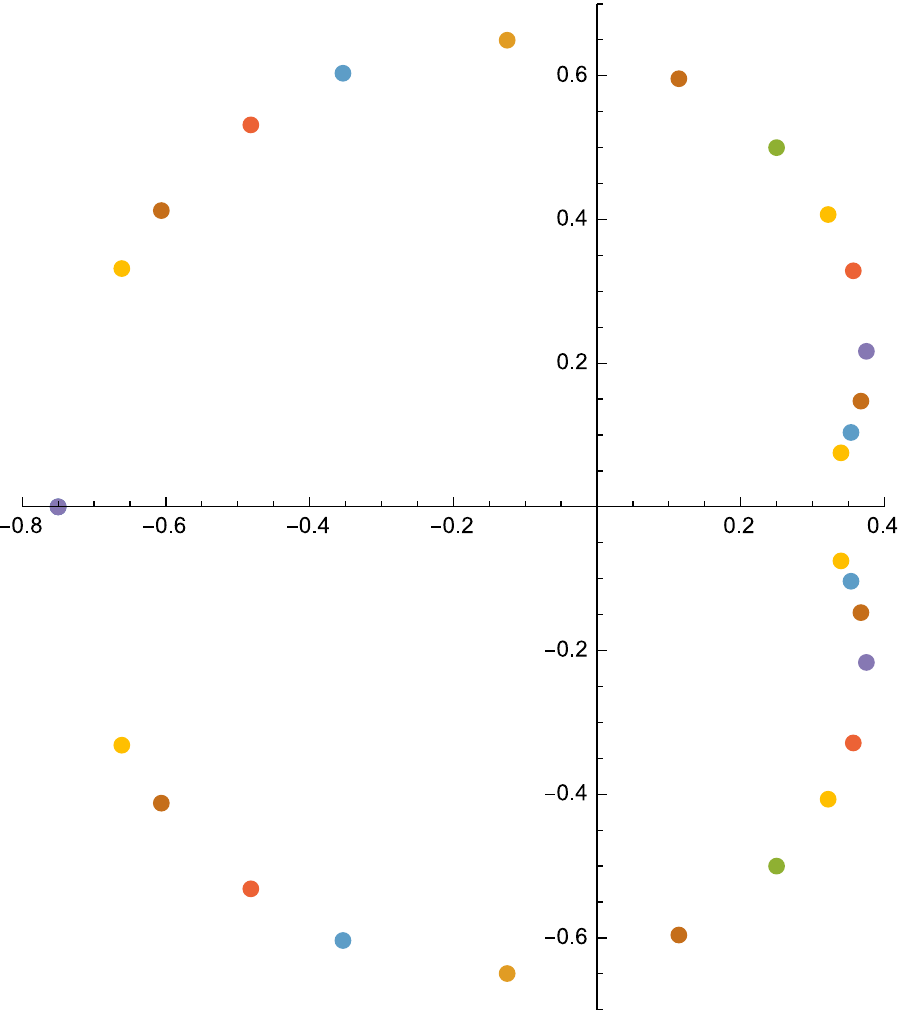}
}
\vspace{-3.5cm}
\be
\label{Fig2}
\ee
\vspace{1.6cm}

\bigskip

This description motivates the following definition:

\bigskip

{\bf
A sequence of polynomials $P_n(x|c)$ has a {\it Mandelbrot property},
if the zeroes of resultants $R_{n,m}(c) = resultants_x(P_n,P_m)$ in the space ${\cal M}$ of moduli $c$
for all $m$ belong to a variety $L_n$ of non-vanishing real codimension in ${\cal M}$.
}

\bigskip

In fact, Mandelbrot property is very sensitive to details.
If, for example, we choose a wrong definition of $F_n = f^{\circ n}(x)-\alpha x$
with $\alpha\neq 1$, no smooth curves emerge: for, say, $\alpha=0$ (green)
and $\alpha=2$ (black) we get

\bigskip

\centerline{
\includegraphics[scale=0.5]{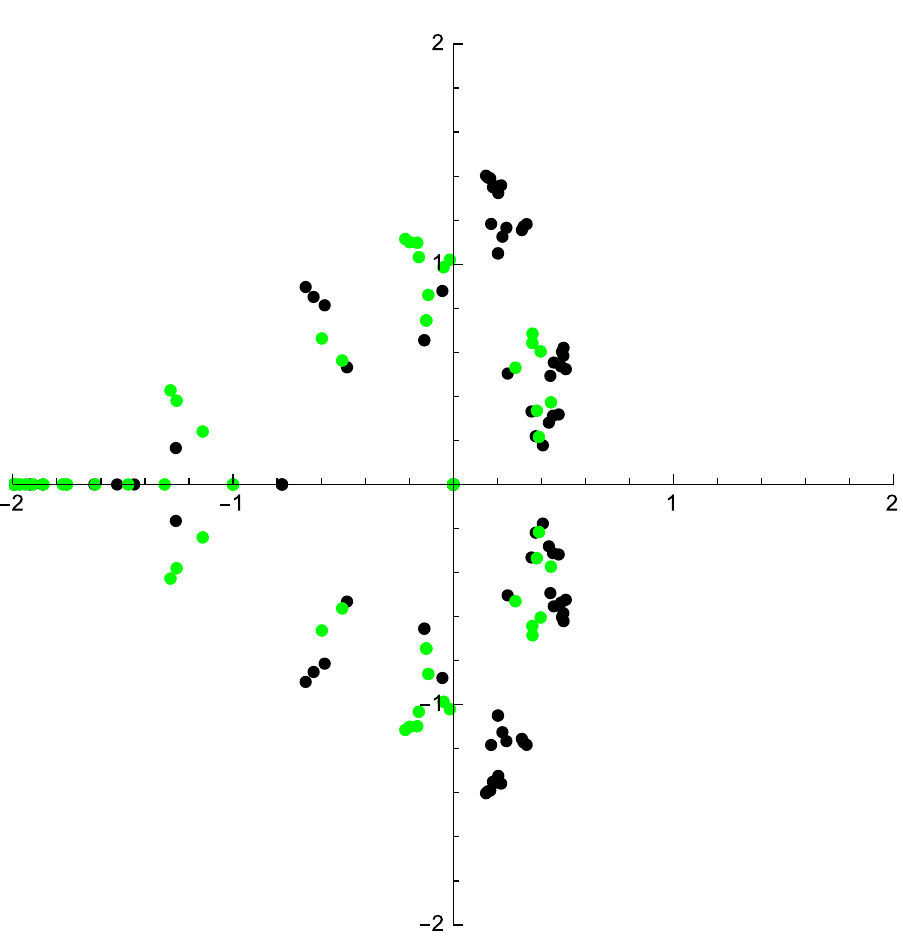}
}
\vspace{-3.5cm}
\be
\label{Fig3}
\ee
\vspace{1.6cm}

\bigskip

\noindent
Thus Mandelbrot property is really distinguished:
it is true for the right objects -- but is easily violated for the wrong ones.

\bigskip

As argued in \cite{DMand},
Mandelbrot property is universally true for sequences of shifted iterated maps $F_n$ -- whatever
is the variety of functions (maps) $f_c(x)$.
The natural question is if there are other interesting examples.

\bigskip

What we demonstrate in the present paper, is that {\bf the sequence of symmetrically colored
HOMFLY polynomials $H_n(q)$} has a good chance to {\bf possess Mandelbrot property}
-- for arbitrary knot, even if the role
of moduli is played by the second HOMFLY variable $A$.

\bigskip

First of all, this is almost {\it a fact}, which we try to make pictorially-obvious in sec.\ref{evid}.
We then discuss a more theoretical evidence in secs.\ref{theofirst}-\ref{Julia}.

\bigskip

Second, one can wonder, if this {\bf fact} could be {\it expected} -- or, in other words,
if there is anything in common between iterated maps and colored HOMFLY, i.e. between
iterations and colorings.
In fact, there is.

If the map $x\longrightarrow f(x)$ was just linear, $x\longrightarrow f\cdot x$
(this is quite an interesting example, if $x$ is a vector, and $f$ -- a matrix),
then iteration is just a matrix product: $f^{\circ n}\sim f^n$, and the latter can be embedded
into a tensor product $f^{\otimes n}$.
Generalizing (or, perhaps, speculating) -- even for non-linear maps one can consider iterations
as some sophisticated projection functor
from the tensor product:
\be
f^{\circ n} = {\cal P}_{ite} \Big(f^{\otimes n}\Big)
\ee
One can now speculate that Mandelbrot property is the one of the projector --
and one can ask if there are other projectors, besides ${\cal P}_{ite}$ which possess this property.

In knot theory cabling method \cite{cabling} allows one to consider colored HOMFLY as certain
projection acting on the $n$-th "power" of the knot ${\cal K}$:
\be
H_n^{\cal K} = {\cal P}_{col} \Big({\cal K}^{\sqcup n}\Big)
\ee
Our claim in this paper is that ${\cal P}_{col}$ also has (a good chance to possess) the Mandelbrot property.

\bigskip

Third, why $A$ is considered as a modulus?
Of course, it would be more natural to look at the moduli space of knots.
What we do in this paper is just the search "under the lamp".
Due to numerous recent advances \cite{IMMMfe,RTmod},
dependence of symmetrically-colored HOMFLY on $A$ is nowadays described in sufficiently
clear and explicit way to make computer experiments and even some analytical studies.
This is not yet so easy for ${\cal K}$-dependence -- though a way is now open by the
suggestions in \cite{mmmrs,mm7}.
However, if one accepts the concept of the Universal Mandelbrot set \cite{DMand},
concrete choice of a single modulus (slicing) to look at is not too essential:
Mandelbrot property will show up in each particular slicing -- and from this
point of view the perhaps-arbitrary choice of $A$ is not a problem.

\bigskip

Forth -- what are the implications of our conjecture?
We prefer to refrain from further speculations at this stage -- despite there can be many.
It is already sufficient that too well-developed branches of theory are now connected,
and the exchange of ideas and concepts can now take place.
Moreover, already the simplest examples we look at below, demonstrate that
resultants know a lot about other non-trivial properties of knot invariants
-- like defects \cite{defect} of differential expansions \cite{diffexpan}
-- what supports our belief that the relation to Mandelbrot world is a deep property
and not just an amusing observation or a joke.
This means that further work is needed to appreciate the real significance
of this newly-emerging research direction, which finally extends the still-mysterious
Mandelbrot property beyond a single example.
Time is needed to put the right accents and provide relevant interpretations.

\section{Technicalities}

We remind that resultant is the function of coefficients of the system of homogeneous non-linear
equations, which vanishes whenever the system has non-vanishing solution:
\be
\vec F(\vec x) = 0, \ \ \ \vec x \neq 0 \ \ \ \ \ \Leftrightarrow \ \ \ \ \
\text{resultant}_{\vec x}(\vec F)\neq 0
\ee
Resultants -- and discriminants, their close relatives,-- are non-linear analogues of
determinants. If known, they make the theory of non-linear equations as transparent
as it is for linear ones -- and the study of resultants and discriminants is the  main
topic in non-linear algebra \cite{GKZ,NLA,ShaM}.
The present knowledge, however, remains restricted --
and the problem attracts much less attention and effort than it deserves.

Fortunately, in the case of a single variable $x$ (i.e. two variables in homogeneous formulation),
which we deal with in the present paper, the story is elementary -- see any textbook on algebra
for explicit formulas, e.g. sec.3.3.3 of \cite{NLA}.
Discriminant of a polynomial $P(x) = \sum_{i=0}^n a_ix^i = \prod_i (x-x_i)$ is equal to
\be
\text{discriminant}_x (P) = \prod_{i<j}(x_i-x_j)^2
\ee
and it is a polynomial in the coefficients $a_i$.
Similarly, a resultant if two polynomials $P_1(x)=\prod_i (x-x^{(1)}_i)$
and $P_2(x)=\prod_j (x-x^{(2)}_j)$,
\be
\text{resultant}_x(P_1,P_2) \sim \prod_{i,j} (x^{(1)}_i-x^{(2)}_j)
\ee
is a polynomial in $a^{(1)}_i$ and $a^{(2)}_j$,
equal to determinant
\be
\text{resultant}_x(P_1,P_2) = {\rm det}\left(\begin{array}{ccccccc}
a_0^{(1)} & a_1^{(1)} & a_2^{(1)} & a_3^{(1)} &\ldots & 0 & 0\\
0 & a_0^{(1)} & a_1^{(1)} & a_2^{(1)} & \ldots & 0 & 0 \\
0 & 0 & a_0^{(1)} & a_1^{(1)}  & \ldots & 0 & 0 \\
&&&\ldots&&\ldots& \\
0 & 0 & 0 & 0 & \ldots &  a_{n_{1}}^{(1)} & 0 \\
0 & 0 & 0 & 0 & \ldots & a_{n_{1}-1}^{(1)} &a_{n_{1}}^{(1)} \\
a_0^{(2)} & a_1^{(2)} & a_2^{(2)} & a_3^{(2)} &\ldots & 0 & 0\\
0 & a_0^{(2)} & a_1^{(2)} & a_2^{(2)} & \ldots & 0 & 0 \\
0 & 0 & a_0^{(2)} & a_1^{(2)}  & \ldots & 0 & 0 \\
&&&\ldots&&\ldots& \\
0 & 0 & 0 & 0 & \ldots &  a_{n_{2}}^{(2)} & 0 \\
0 & 0 & 0 & 0 & \ldots & a_{n_{2}-1}^{(2)} &a_{n_{2}}^{(2)}
\end{array}\right)
\ee
and matrix has the size $n_{1}+n_{2}$.

\bigskip

The second ingredient of our consideration --
colored HOMFLY polynomials -- can be taken from the database in \cite{knotebook}:
they are result of various calculations by a variety of modern methods, see
\cite{RTmod}-\cite{mm7} and references therein.
HOMFLY in topological framing are actually Laurent polynomials, and to apply
standard definitions of resultants, they should be multiplied by a certain power of $q$
to become ordinary polynomials -- what is always assumed in this paper.

\section{Pictorial evidence
\label{evid}}

With this data we can easily plot zeroes or colored HOMFLY resultants.

\bigskip

\noindent
Zeroes of
${resultant}_{q^2}(H_1^{4_1}, H_k^{4_1})$
for the figure eight knot ${\cal K} = 4_1$:

\centerline{
\includegraphics[scale=0.5]{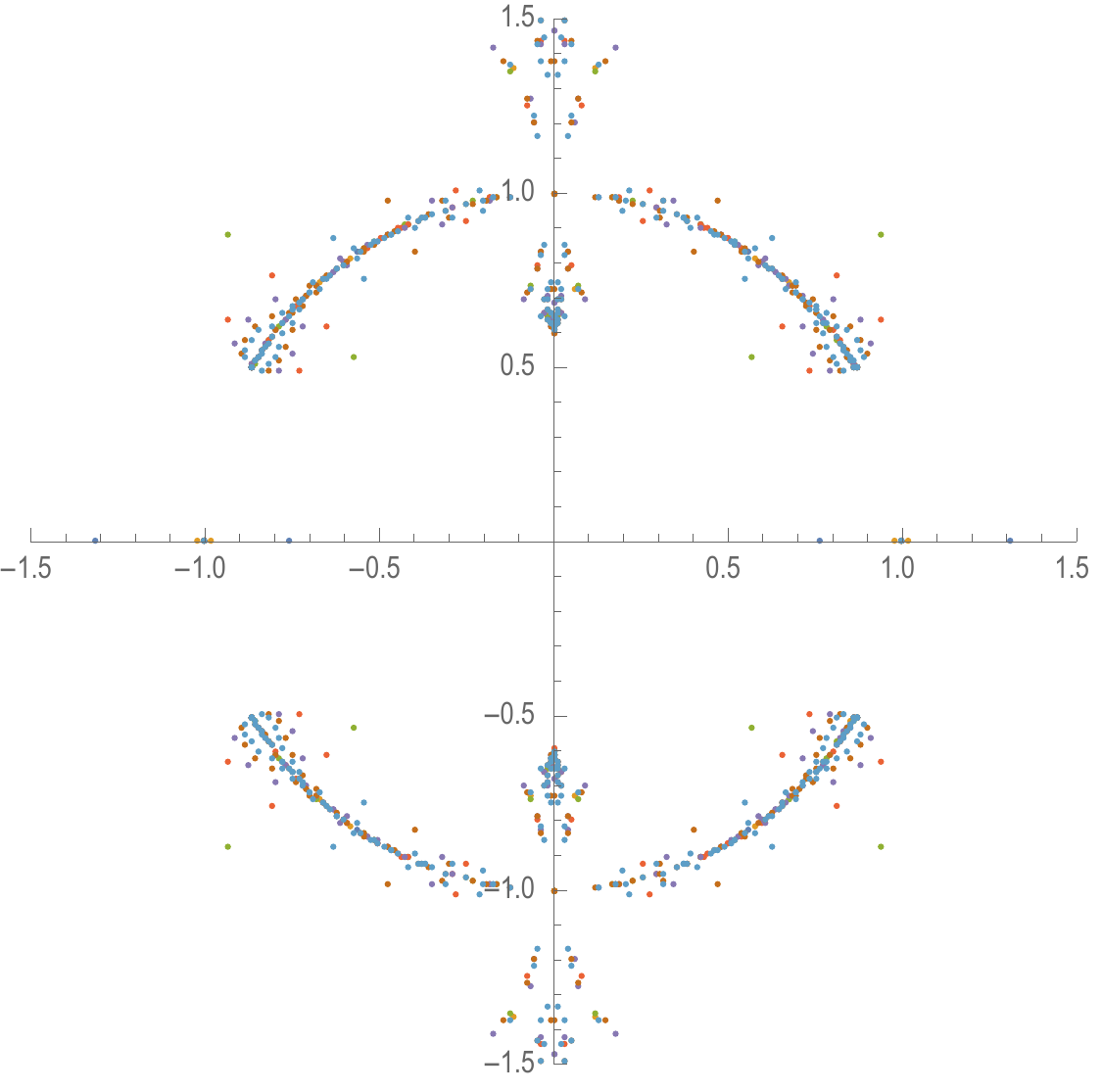}
}
\vspace{-3.5cm}
\be
\label{Fig4}
\ee
\vspace{1.6cm}

\noindent
and those of
${resultant}_{q^2} (H^{4_1}_2, H^{4_1}_k)$:

\centerline{
\includegraphics[scale=0.5]{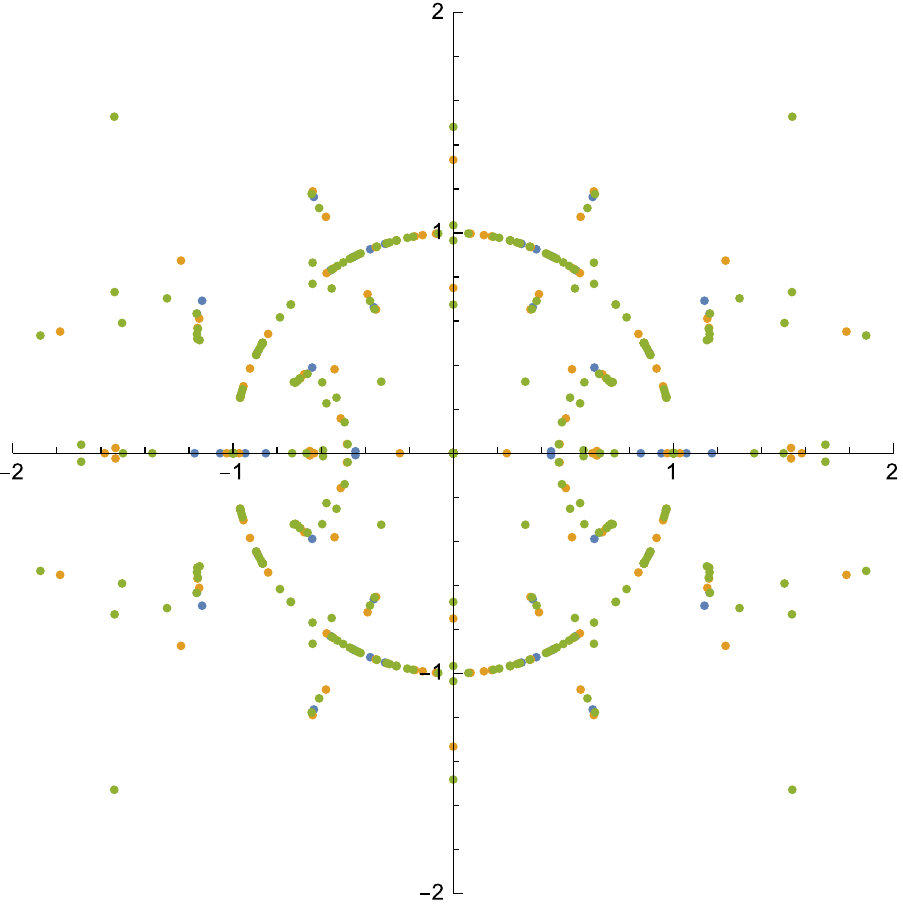}
}
\vspace{-3.5cm}
\be
\label{Fig5}
\ee
\vspace{1.6cm}

\noindent
We see that instead of being chaotically scattered all over the complex $A$-plane,
these zeroes form certain one-dimensional structures, i.e. exhibit the {\it Mandelbrot property}.
Note, that the two segments of the unit circle in the first picture
do not form a closed curve and can not serve as a {\it boundary}
of any 2-dimensional domain -- as it was in the Mandelbrot case.
Still, the second picture extends these segments to a bigger fraction of the  unit circle,
and zeroes of {\it all} the resultants $R_{k,l}$ seem to populate densely the entire closed curve.

\bigskip

Situation for more complicated knots can look less convincing:
the same plots for the defect-2 knot ${\cal K} = 7_1$

\bigskip

\centerline{
\includegraphics[scale=0.5]{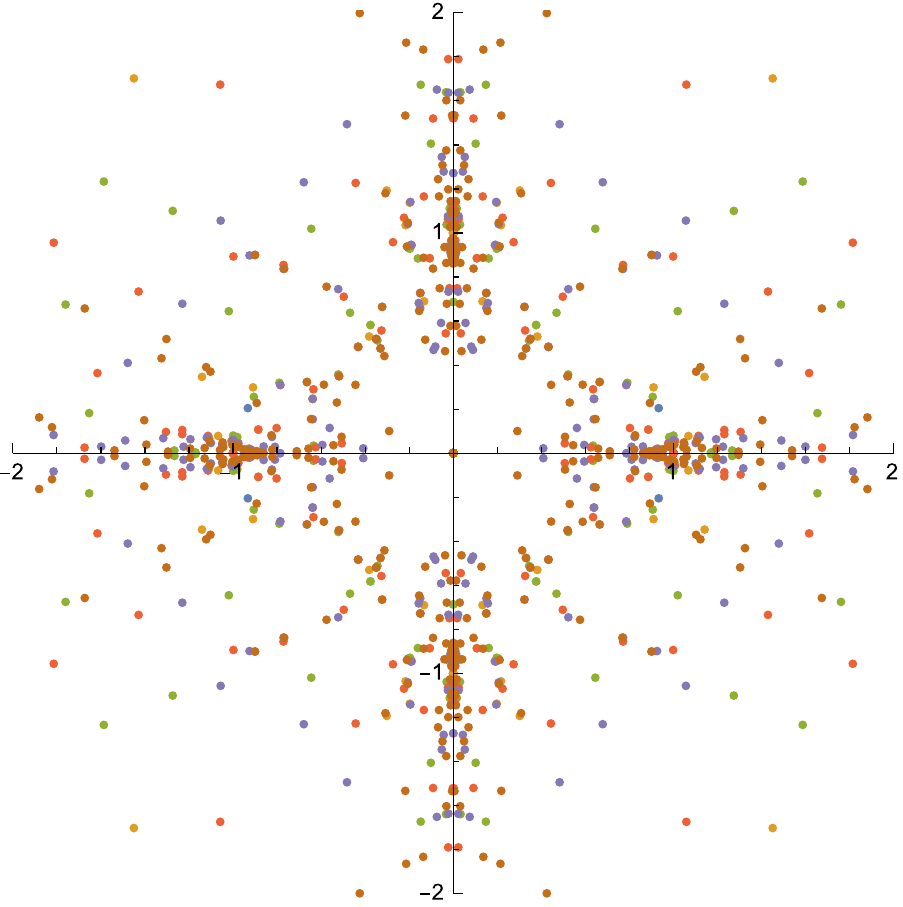} \ \ \ \ \ \ \ \ \ \ \ \ \ \ \
\includegraphics[scale=0.5]{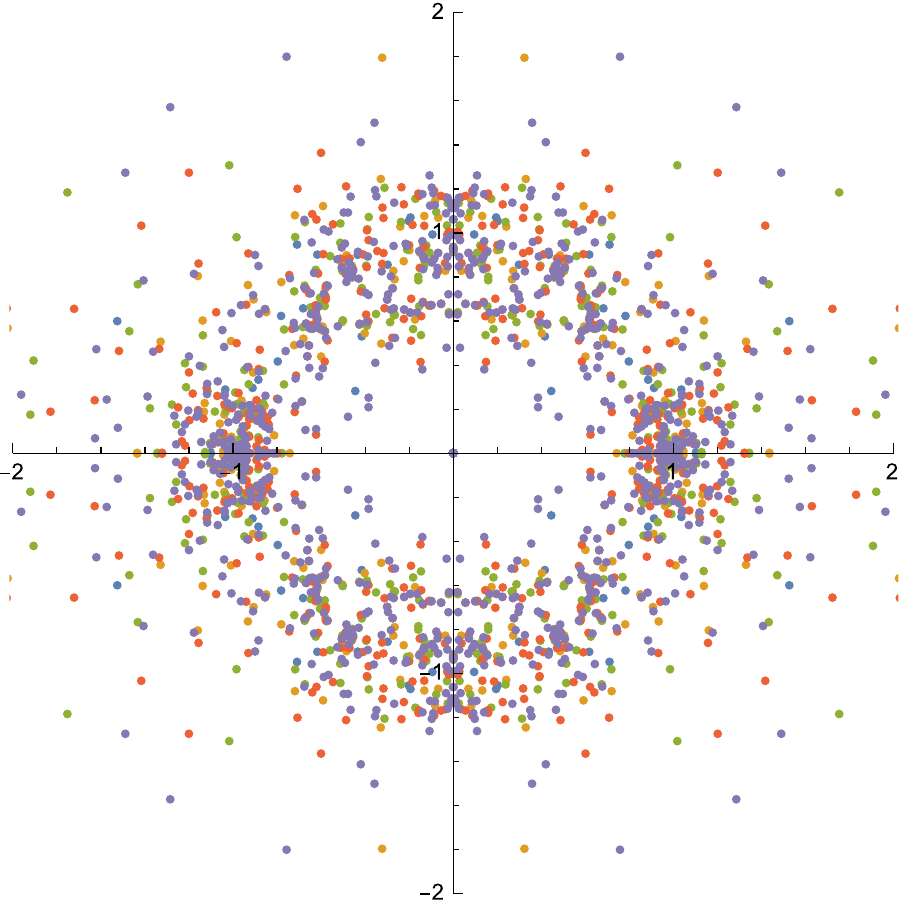}
}
\vspace{-3.5cm}
\be
\label{Fig6}
\ee
\vspace{1.6cm}

\bigskip

\noindent
look more like full 2-dimensional patterns.
However, this is only because of the poor resolution.
Looking closer, we see, for example:

\bigskip

\centerline{
\includegraphics[scale=0.5]{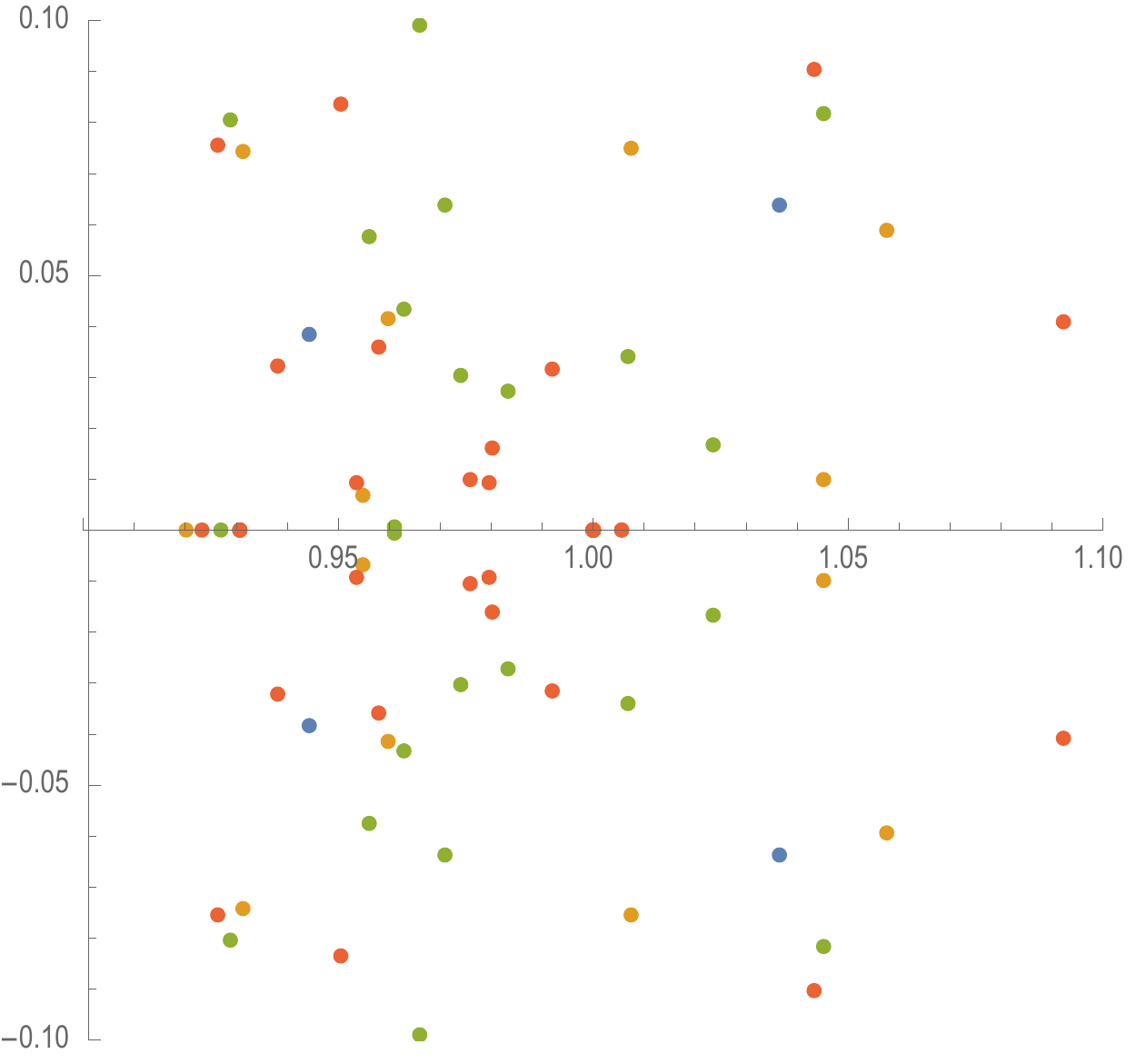}
}
\vspace{-4cm}
\be
\label{Fig7}
\ee
\vspace{2.1cm}

\bigskip

\noindent
For more details and examples see \cite{knotebookRES}.

\section{Meditation: how special are sequences with Mandelbrot property?}

While pictures in the previous section obviously reveal {\it some} correlation between
zeroes of different resultants of colored HOMFLY polynomials,
they are rather far from the clarity of (\ref{Fig1}) and (\ref{Fig2}) and can hardly
be treated as an experimental  {\it proof} the Mandelbrot property.
At best they resemble (\ref{Fig3}) rather than (\ref{Fig2}), what can imply that HOMFLY
polynomials deserve to be properly shifted to improve the situation.
In the case of iterated maps there was a distinguished shift $-x$, dictated
either by the formulation of the problem (search for closed orbits)
or, by an algebraic criterium of divisibility of $F_{kn}$ by $F_n$,
which a sort of "ties" at least some zeroes of $F_{kn}$ to those of $F_n$
and helps to choose the proper resultants.
For HOMFLY the story is different:
there is no clear way to make a shift, which would make $H_{kn}$ divisible by $H_n$.
Still, {\it some} naive shifts improve the pattern, though not drastically enough.
For example, a natural option is to shift $H_n$ by one, and here is what we get:

\bigskip

roots of $resultant_{q^2} \left(
\frac{H_2^{3_1}-1}{\{A/q\}}, \frac{H_j^{3_1} - 1}{\{A/q\}}
\right):$

\bigskip

\centerline{
\includegraphics[scale=0.5]{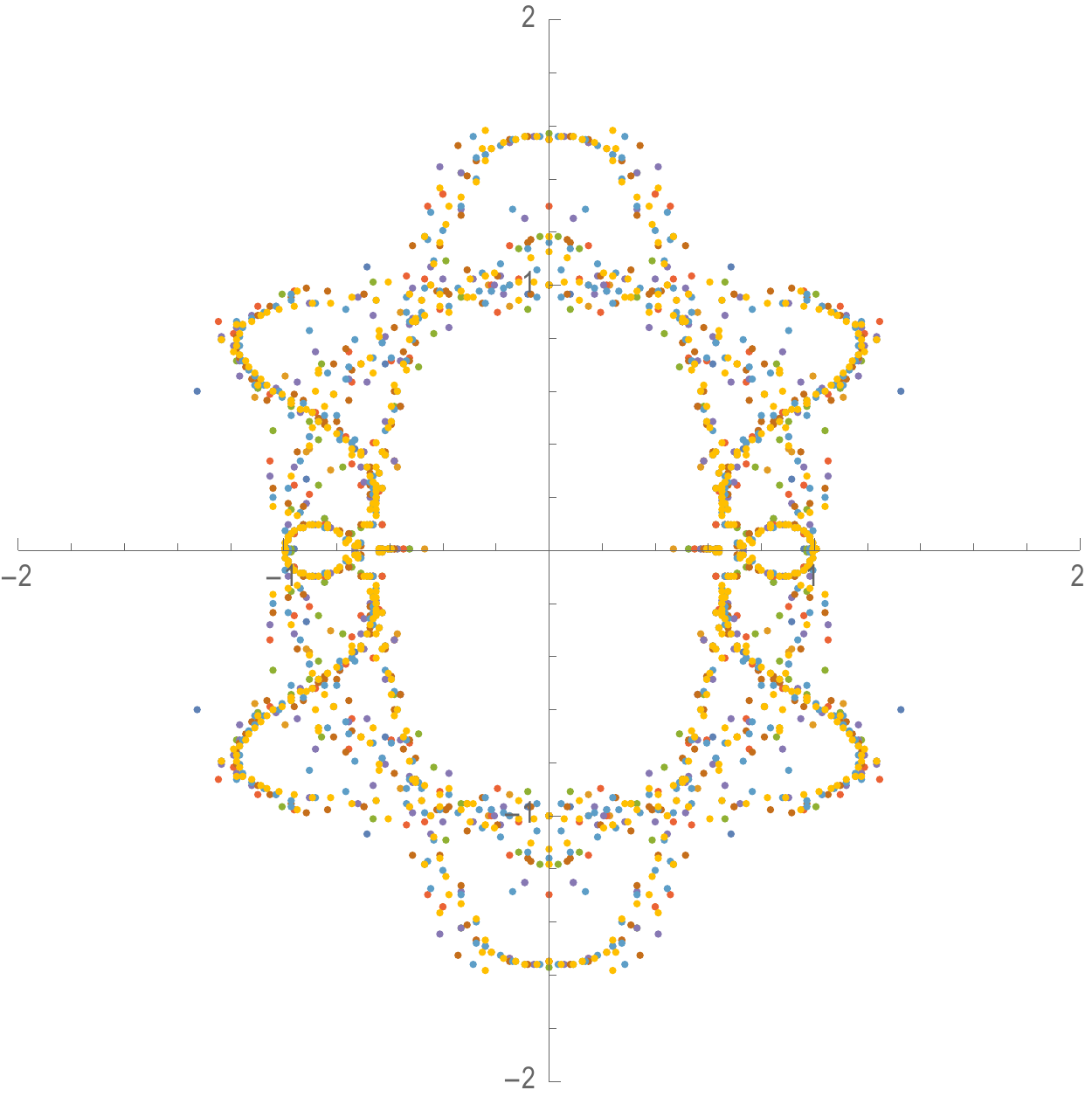}
}
\vspace{-4.4cm}
\be
\label{Fig8}
\ee
\vspace{2.5cm}

\bigskip

roots of $resultant_{q^2} \left(
\frac{H_2^{3_1}-H_1^{3_1}}{\{A/q\}}, \frac{H_j^{3_1} - H_1^{3_1}}{\{A/q\}}
\right):$

\bigskip

\centerline{
\includegraphics[scale=0.5]{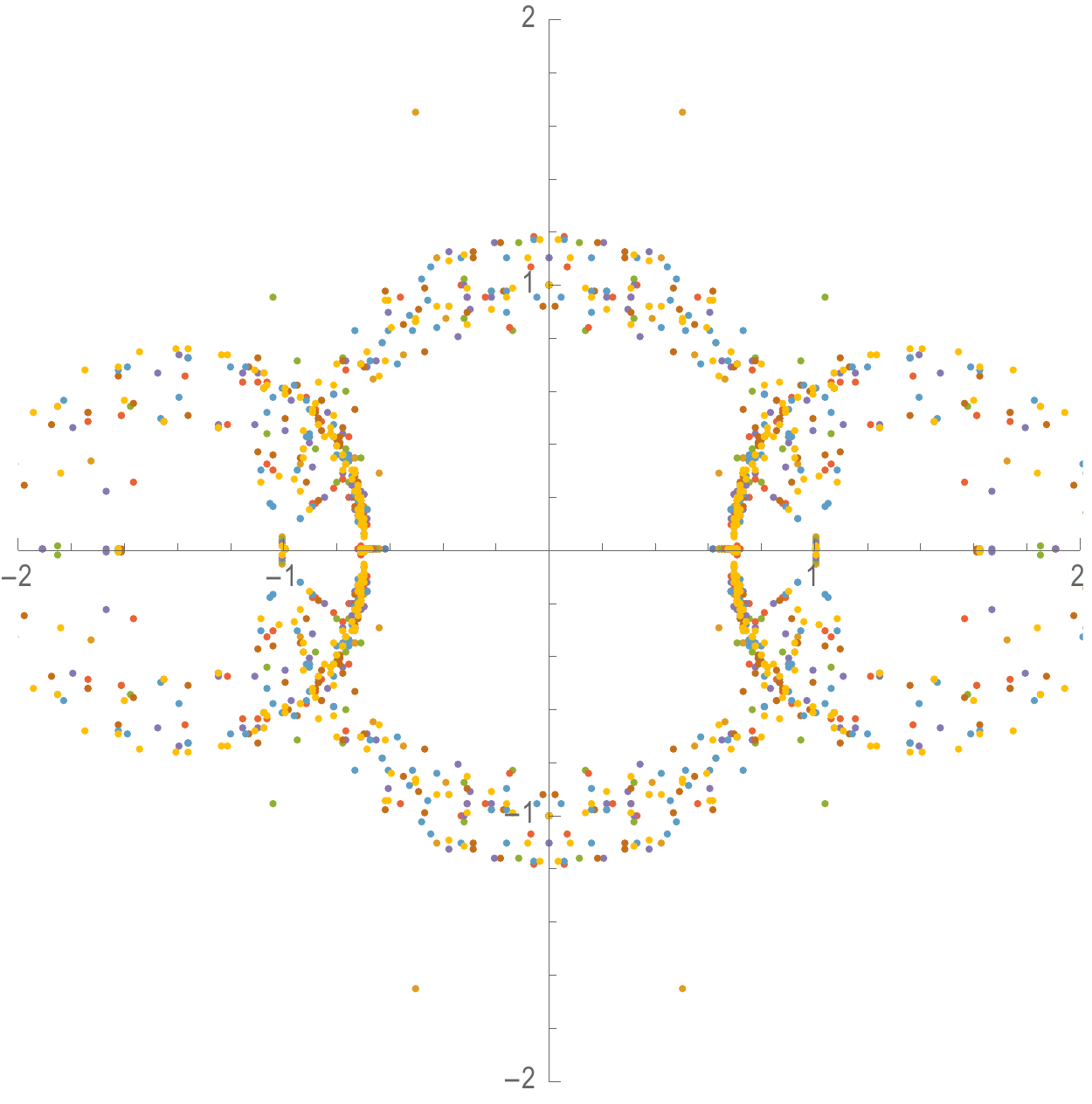}
}
\vspace{-4.2cm}
\be
\label{Fig9}
\ee
\vspace{2.3cm}

\bigskip

One can now ask a natural question --
if these somewhat obscure pictures manifest anything at all:
perhaps, patterns of this quality are provided by just {\it arbitrary}
sequence of polynomials.
To illustrate the actual state of affairs, we consider {\it not an arbitrary},
but a very special sequence $\{G_n (x,z)\}$,
provided by the $t$-expansion of the generating function
\be
\exp\left(\sum_{k=1}^m t^k P_k(x,z)\right) = \sum_{n=0}^\infty t^n G_n(x,z)
\ee
i.e. $G_n$ are symmetric Schur functions of the polynomial-valued time-variables $P_k$
-- and plot the zeroes of $resultant_x(G_k,G_l)$ in the complex-$z$ plane.

At $m=1$ all $G_n\sim P_1^n$ and all resultants are just vanishing.

At $m=2$ they all reduce to a single $resultant_x(P_1,P_2)$
and there is nothing interesting.

However, starting from $m=3$, the situation changes:
e.g. for
\be
P_1 = x + z \nn\\
P_2 = x^2 + z^2 \nn\\
P_3 = x^3 z^2
\label{Exa1}
\ee
the pattern of resultant zeroes looks as follows:

\centerline{
\includegraphics[scale=0.5]{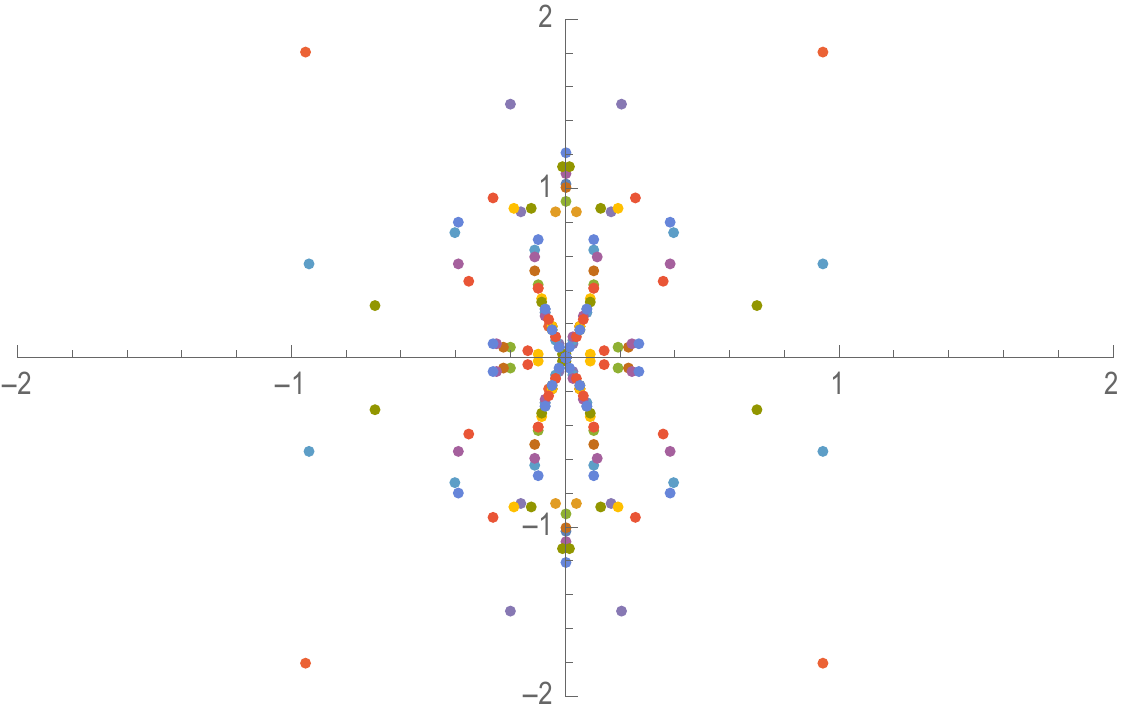}
}
\vspace{-3.5cm}
\be
\label{Fig10}
\ee
\vspace{1.6cm}


\noindent
while for
\be
P_1 = z^3 + z^2 -  x^2z \nn\\
P_2 = z^5 - x^7z^3  \nn\\
P_3 = x^3z^2  + x^2 z^4
\label{Exa2}
\ee
it becomes

\vspace{0.5cm}
\be
\label{Fig11}
\ee
\vspace{-2.6cm}

\centerline{
\includegraphics[scale=0.5]{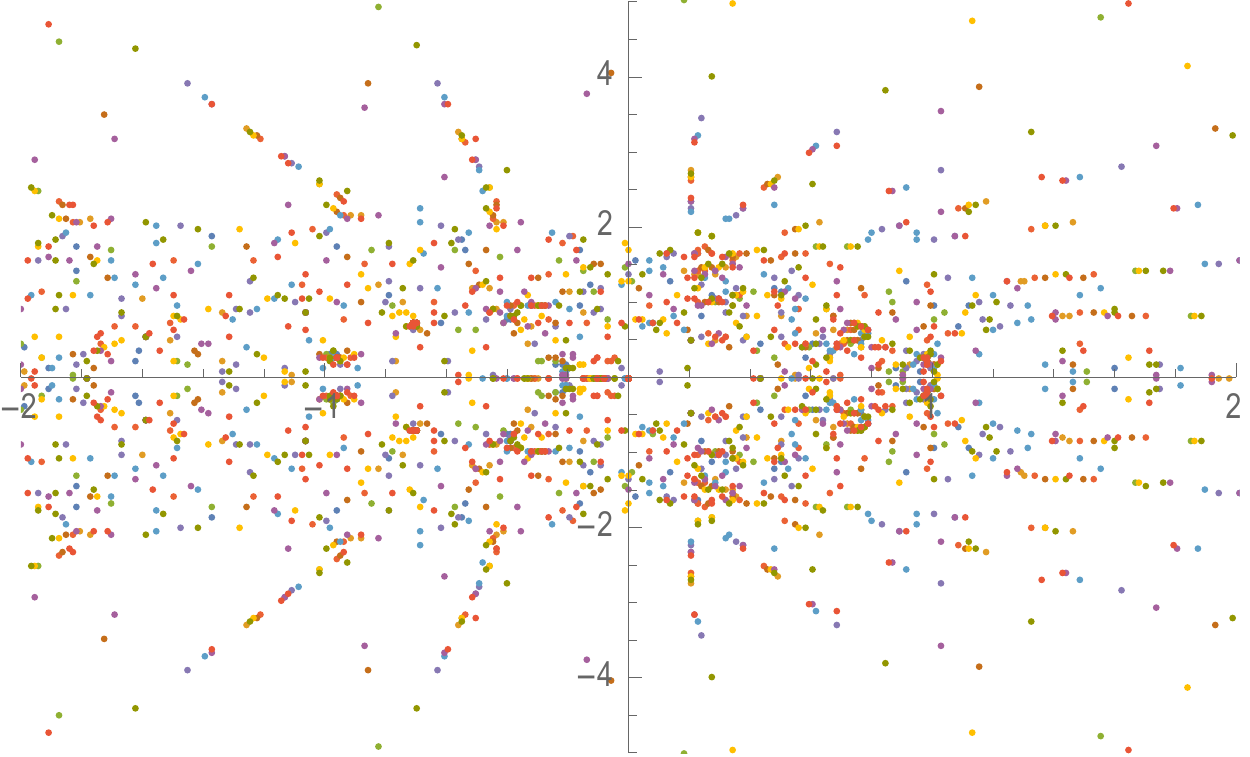}
}


We clearly see that with increase of $m$ and/or of the complexity of the polynomials $P_k$
the one-dimensional structures quickly dissolve in two-dimensional clouds
and Mandelbrot property, if at all present, gets destroyed --
and this happens already for {\it a very special} sequence, provided by a Schur transform
from just a few simple polynomials.

In fact, for $m=3$ a zero of
$resultant_x(G_m,G_n)$ is associated with a common solution of two equations
\be
\left\{\begin{array}{c} P_2=a_{mn}\cdot P_1^2 \\ P_3=b_{mn}\cdot P_1^3 \end{array}\right.
\ee
i.e.
\be
resultant_x(G_m,G_n) \ \sim \ resultant_x\Big( P_2-a_{mn}\cdot P_1^2,\  P_3-b_{mn}\cdot P_1^3\Big)
\label{resres}
\ee
where $a$ and $b$ are numerical constants, depending on $m$ and $n$  -- and it
is the diversity of the coefficients $a_{mn}$ and $b_{mn}$ which is responsible
for the {\it lack} of Mandelbrot property.
A detailed study of this simple example is of its own interest, but it would take us too far
away from the main subject of the present paper.

\bigskip

Instead we now return to HOMFLY resultants and proceed from pictures of sec.3 to formulas.
We do not do much in the present paper, but the kind of questions
to address should be clear from the simple examples below.

\section{Fundamental HOMFLY and HOMFLY for other transposition-\-sym\-metric diagrams
\label{theofirst}}

Colored HOMFLY polynomials are defined for arbitrary representation (Young diagram) $R$
and they possess the important symmetry: transposition of $R$ is equivalent to inversion
of $q^2$:
\be
H_{\tilde R}(q|A) = H_R\left(-\frac{1}{q}\,\Big|A\right)
\ee
In result for transposition-symmetric diagrams, $\tilde R=R$, HOMFLY actually depend on
$q$ via the square of the variable $z=\{q\}\equiv q-1/q$.
This has an immediate consequence for its discriminant.
By definition, ${discriminant}_{q^2}H$ vanishes whenever the system
\be
\left\{ \begin{array}{c} H=0, \\
\frac{\p H}{\p q^2}=0
\end{array}\right.
\ee
has a solution -- and since $q^2\frac{\p H}{\p q^2} = \frac{q^4-1}{q^2}\cdot\frac{\p H}{\p z^2}$,
we see that $H(q=\sqrt{1}|A)=0$ is always a solution.
This means that
\be
{\rm for} \ \tilde R=R \ \ \ \ \ \
{discriminant}_{q^2}H_R \ \sim \ H_R(q=1|A)\cdot H_R(q=i|R)
\label{factrasymR}
\ee
This fact can be reinterpreted as follows:
if considered as a polynomial with real coefficients, $H_R$
with $\tilde R=R$ has distinguished
set of zeroes, lying at the real axis $z^2$.
From the point of view of $q^2$ this real axis is further separated into three regions:
$z^2<-2$,  $z^2>2$ and $-2<z^2<2$, where $q^2$ lies respectively on negative and positive rays
$q^2<-1$, $q^2>1$ and unit circle $|q|=1$ (comp.with \cite{Hunicircle}).
Clearly, discriminant  $discriminant_{q^2}(H_{[1]})$ should have zero,corresponding to the
bifurcation points $q^2=\pm 1$.

Looking at examples, we observe that (\ref{factrasymR}) is always true, at least for representations
$R=[1], [21], [22]$, where many results are available.
Moreover, for the fundamental representation $R=\Box=[1]$
the l.h.s. {\it equals} the r.h.s. for all knots with
the vanishing defect \cite{defect} $\delta^{\cal K}=0$:
\be
\delta^{\cal K}=0 \ \ \ \ \Longrightarrow \ \ \ \
D^{\cal K}_{[1]}(A) \equiv \sqrt{\frac{
discriminant_{q^2}H_{[1]}^{\cal K}}{H_{[1]}^{\cal K}(A|q=1)\cdot H_{[1]}^{\cal K}(A|q=i)}} = 1
\ee
We introduced the square root into the definition of $D^{\cal K}$, because for knots
with non-vanishing defect the ratio is a full square(!)
--  and this seems also true for other symmetric Young diagrams $R$.

Moreover, it factorizes further.
In the simplest case of $5_1$ with defect one,
$D^{5_1} = H_{[1]}^{\cal K}(A|q=\xi)\cdot H_{[1]}^{\cal K}(A|q=i\xi)$
with $\xi^2=\frac{1\pm\sqrt{5}}{2}\cdot i\ $ pure imaginary,
i.e. $\xi \sim (1+i)\cdot (\text{a real number})\ $
-- however, such simple description of emerging factors fails for more complicated knots.

\section{More on the role of the special polynomials}

One of the two polynomials, which appeared as a factor in discriminant,
is the well known {\it special polynomial} $H_R(q=1|A)$.
We remind that the special polynomials at $q^2=1$ are always factorized
\be
H_R(q=1|A) = H_{[1]}(q=1|A)^{|R|} \ \ \ \ \forall \ R
\label{spepo}
\ee
Likewise Alexander polynomials for the single hook diagrams $R$ satisfy
\be
H_R(q|A=1)=H_{[1]}(q^{|R|}|A=1)\ \ \ \ {\rm for} \ R=[a,1^b]
\label{Alex}
\ee

\bigskip

The property (\ref{spepo}) implies that all $H_R(q|A)$ of a given knot have
common zeroes at $q=1$ and at $A$, which are the zeroes of $H_{[1]}(q=1|A)$,
i.e. that all the resultants
\be
resultant_{q^2}(H_R,H_{R'}) \ \vdots \ H_{[1]}(q=1|A) \ \ \ \ \forall \ R,R'
\ee

In fact the divisibility of resultants and discriminants is richer:
for arbitrary knots
$resultant(H_k, H_l)$ is always divisible by a power of the special polynomial:
\be
resultant(H_k, H_l) \ \vdots \ H_1(q=1|A)^{^{
\left [\frac{kl+1}{2} \right]}}
\ee
Moreover,

\be
\begin{array}{ccc}
r(1,2) &\vdots & H_{1,1} \\
r(1,3)&\vdots & H_{1,1}^2 H_{1,2} \\
r(1,4)&\vdots & H_{1,1}^2 \\
r(1,5)&\vdots & H_{1,1}^3 H_{1,2} \\
r(2,3) &\vdots& H_{1,1}^3 H_{2,2} \\
r(2,4) &\vdots& H_{1,1}^4 H_{2,2} \\
r(2,5) &\vdots& H_{1,1}^5 H_{2,2} \\
r(3,4) &\vdots& H_{1,1}^6 H_{2,2} H_{3,3}^2 \\
r(3,5) &\vdots& H_{1,1}^8 H_{1,2} H_{2,2} H_{3,3}^2 \\
r(4,5) &\vdots& H_{1,1}^{10} H_{2,2}^3 H_{3,3}^2 H_{4,4}^2 \\
\ldots
\end{array}
\ee
and
\be
\begin{array}{ccc}
d_1   &\vdots &     H_{1,1} H_{1,2} \\
d_2   &\vdots&     ----              \\
d_3   &\vdots&     H_{1,1}^2           \\
d_{21}  &\vdots&    H_{1,1}^3 H_{21,2}     \\
d_4   &\vdots&     H_{1,1}^4               \\
d_{31}  &\vdots&    H_{1,1}^4                  \\
d_{22}  &\vdots&    H_{1,1}^4 H_{2,2}            \\
d_{211} &\vdots&   H_{1,1}^4                       \\
d_{1111} &\vdots&  H_{1,1}^4                        \\
\ldots
\end{array}
\ee
where we denoted $H_{Q,j} = H_Q (q^{2j} = 1)$.

\bigskip

From the factorization property we see that
\be
resultant_{q^2} (H_1, H_r) = 0 \ \ \ \ {\rm if} \ \ sp(A)=0
\ee
and
\be
resultant_{q^2} (H_2, H_r) = 0 \ \ \ \ {\rm if} \ \ sp(A)\cdot sp(A^2) =0
\ee
if
and so on.
Actually it turns out that these peculiar zeroes of resultants
are the limiting points of the sequence of
roots. Here is a pictorial and numerical evidence for
$7_1$-knot:  the resultant zeroes  in this case

\bigskip

\centerline{
\includegraphics[scale=0.5]{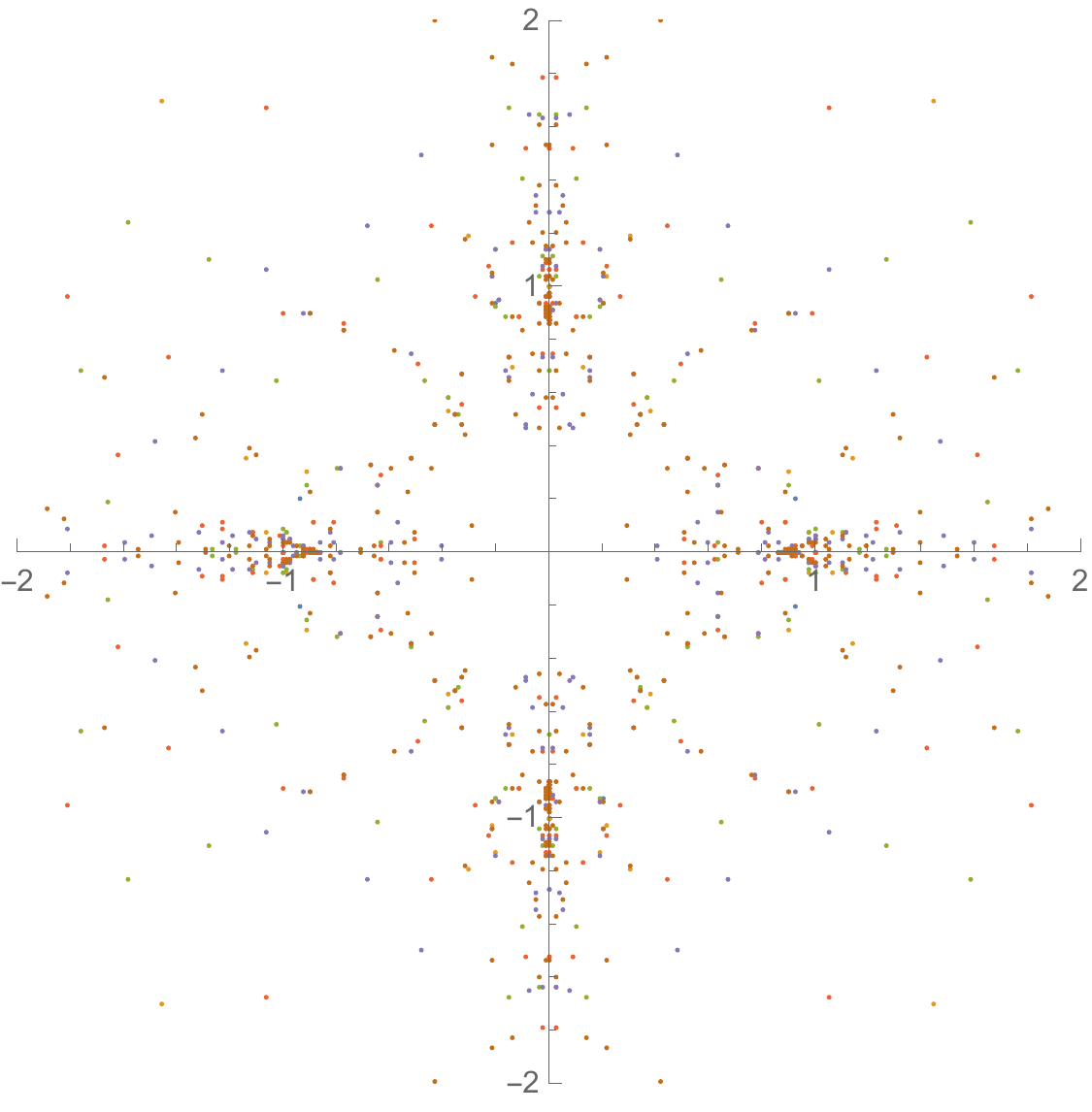}
}
\vspace{-4cm}
\be
\label{Fig13}
\ee
\vspace{2.1cm}

\noindent
are clearly concentrated near
the roots $A = \pm \frac{\sqrt{3}}{2}$ of the special polynomial.
For example, $resultant_{q^2} (H_1, H_7)$ has 414 roots, and 90 of them lie at distances
less then $0.2$ from the roots of $sp(A)$.

Similar is the distribution of $resultant_{q^2}(H_1, H_r)$ for the $7_2$-knot:

\bigskip

\centerline{
\includegraphics[scale=0.5]{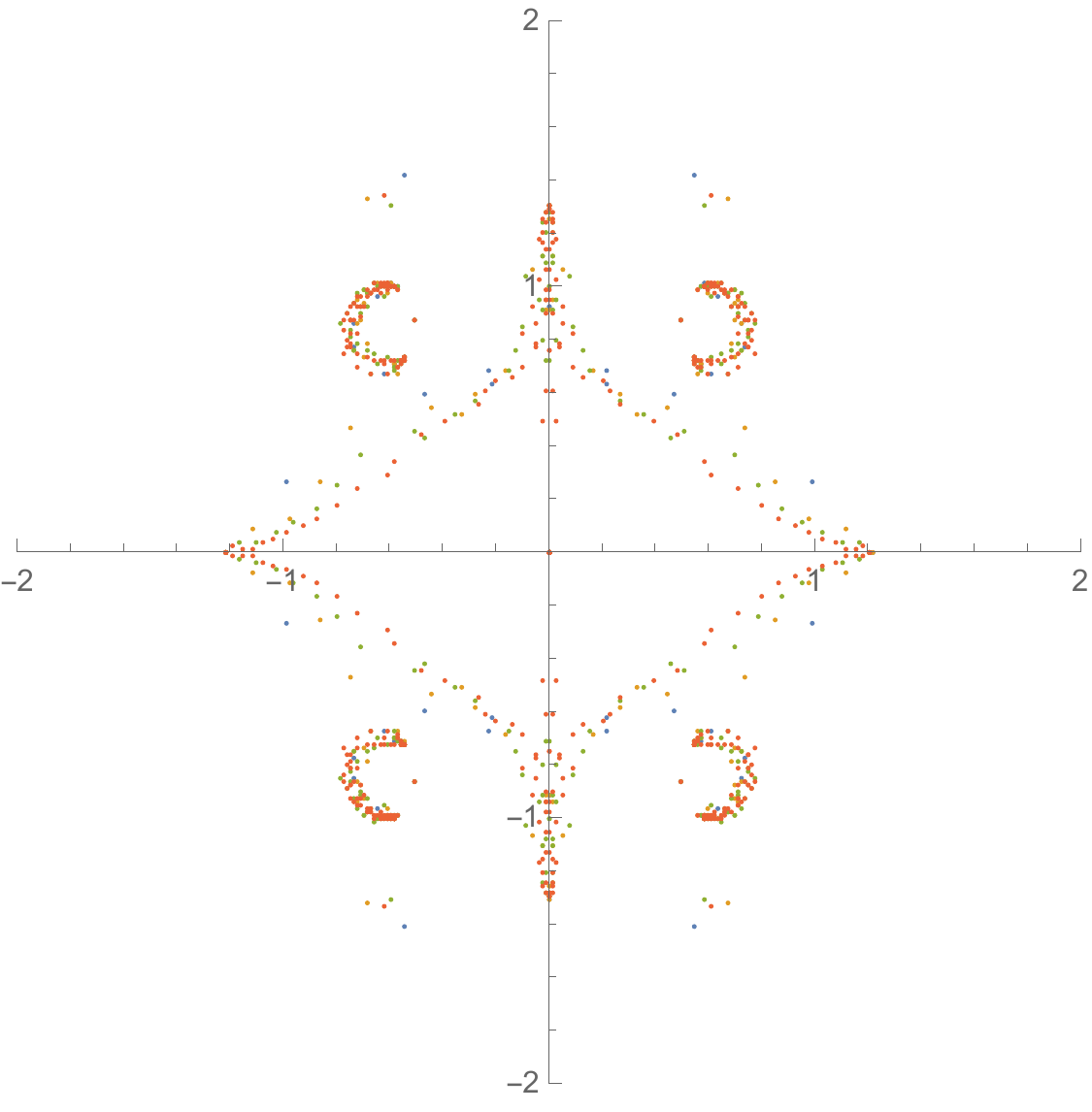} \ \ \ \ \ \ \ \
\includegraphics[scale=0.5]{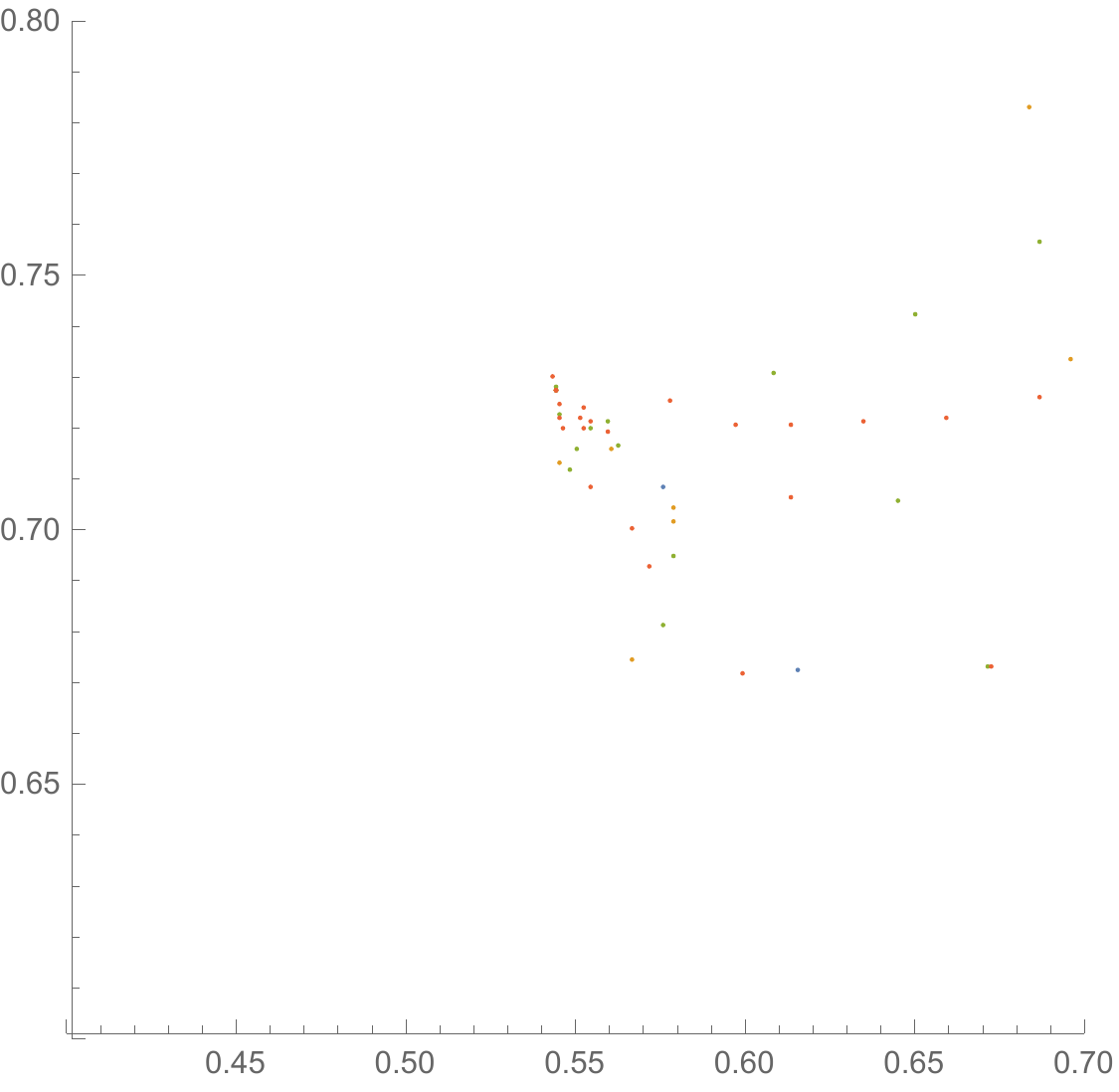}
}
\vspace{-4cm}
\be
\label{Fig14}
\ee
\vspace{2.1cm}

\bigskip

\section{Hopes for divisibility: differential expansion}

As mentioned in the Introduction, the seemingly important property of iterated maps
was that $F_{kn}$ always contains $F_n$ as a factor -- simply because a closed orbit
of order $n$ is always an orbit of order $nk$.
Though obvious, this fact alone requires a correlation to exist between the roots of
$F_{kn}$ and $F_n$ -- and can therefore play a role in occurrence of the Mandelbrot
property.
It is a natural question, if similar divisibility can emerge for colored HOMFLY.

A possible approach to this problem is the study of differential expansion.
Namely, for symmetric-representation HOMFLY of any prime knot ${\cal K}$  we have \cite{defect}:
\be
H_r^{\cal K} = 1 + [r]\{Aq^r\}\{A/q\}\cdot G_1^{\cal K}(q|A) + \sum_{j}^r \frac{[r]!}{[j]![r-j]!}
\{A/q\}\prod_{i=0}^{j-1} \{Aq^{r+i}\} \cdot G_j^{\cal K}(q|A)
\ee
As we just discussed, at $q=1$ all functions $G_j$ are such, that
(\ref{spepo}) is true and divisibility takes place.
However, for $q\neq 1$ this is no longer so.

Of course, all the differences $H_r-1$ have a common zero at $A=q$, but this simply
means that of interest can be the ratios
\be
h_r = \frac{H_r-1}{\{A/q\}}
\ee
Each $h_r$ is divisible by $\{Aq^r\}$ -- and to increase the chances for
common zeroes, it deserves switching to either
$\tilde h_r = h_r(q|Aq^{-r})$ or to $\hat h_r = h_r(q^r|A)$.

We plot zeros for
\be
resultant_{q^2} \left(
\frac{H_i (q^j) - 1}{\{A/q^j\} \{A q^{i j}\}}, \frac{H_j (q^i) - 1} {\{A/q^i\} \{A q^{ij}\}}
\right)
\ee

\bigskip

\centerline{
\includegraphics[scale=0.5]{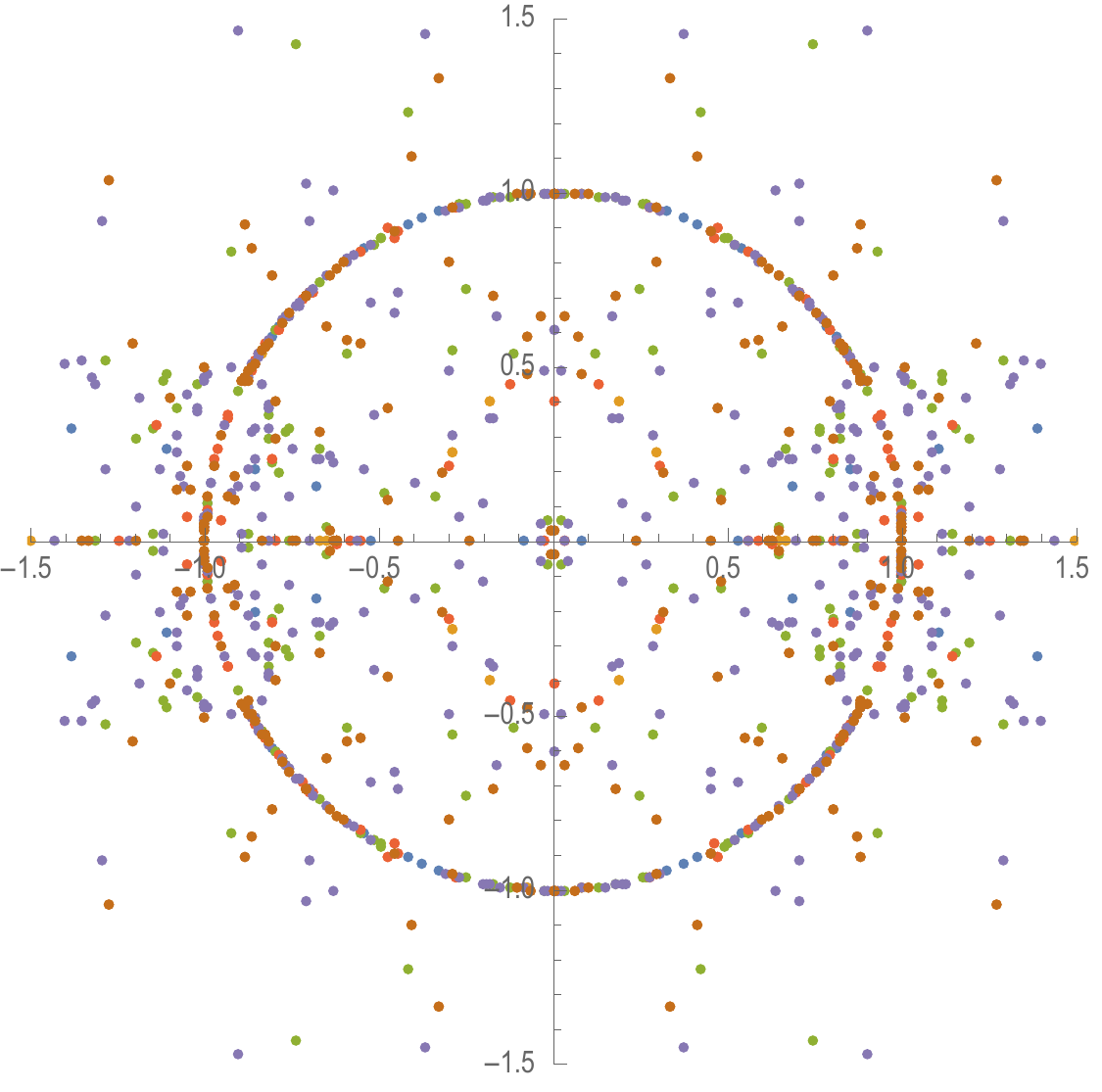}
}
\vspace{-4cm}
\be
\label{Fig13}
\ee
\vspace{2.1cm}

and those for
\be
resultant_{q^2} \left(
\frac{H_i - 1}{\{A/q\} \{A q^{i}\}}, \frac{H_j  - 1} {\{A/q\} \{A q^{j}\}}
\right)
\ee

\bigskip

\centerline{
\includegraphics[scale=0.5]{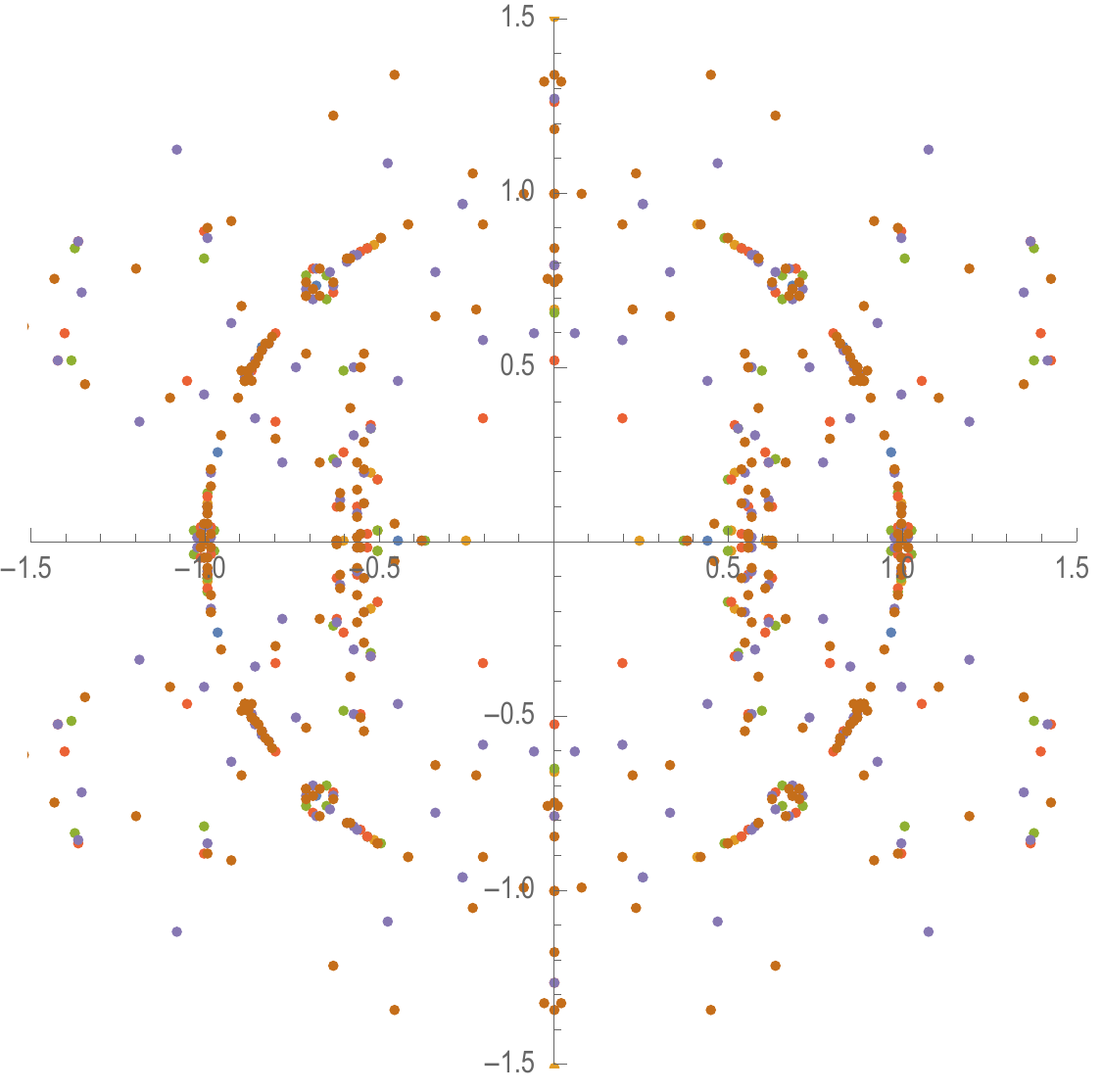}
}
\vspace{-4cm}
\be
\label{Fig13}
\ee
\vspace{2.1cm}

\section{Julia bundle
\label{Julia}}

Over each point in the $A$-plane we can put a pattern of zeroes of the
polynomials $H_n$ in $q^2$.
In the case of Mandelbrot these zeroes of $F_n$
(collection of closed orbits) are called Julia sets, and the entire
construction is a Julia "bundle" over the Mandelbrot set --
we keep the same name in our context.
Julia sets are reshuffled at the lines $L_n$ and it is instructive to
see, how this looks in the case of colored HOMFLY.

We begin with a few illustrative pictures:
plot the $q$-roots of $H_r^{4_1}$ at various values of $A$ for the figure eight knot:

\bigskip

$$
A = 0.01 \ \ \ \ \ \ \ \ \ \ \ \ \ \ \ \ \ \ \ \ \ \ \ \ \ \ \ \ \ \ \ \  \ \ \ \ \ \ \ \ \ \ \ \ \ \
A = 0.4:
$$

\bigskip

\bigskip

\centerline{
\includegraphics[scale=0.5]{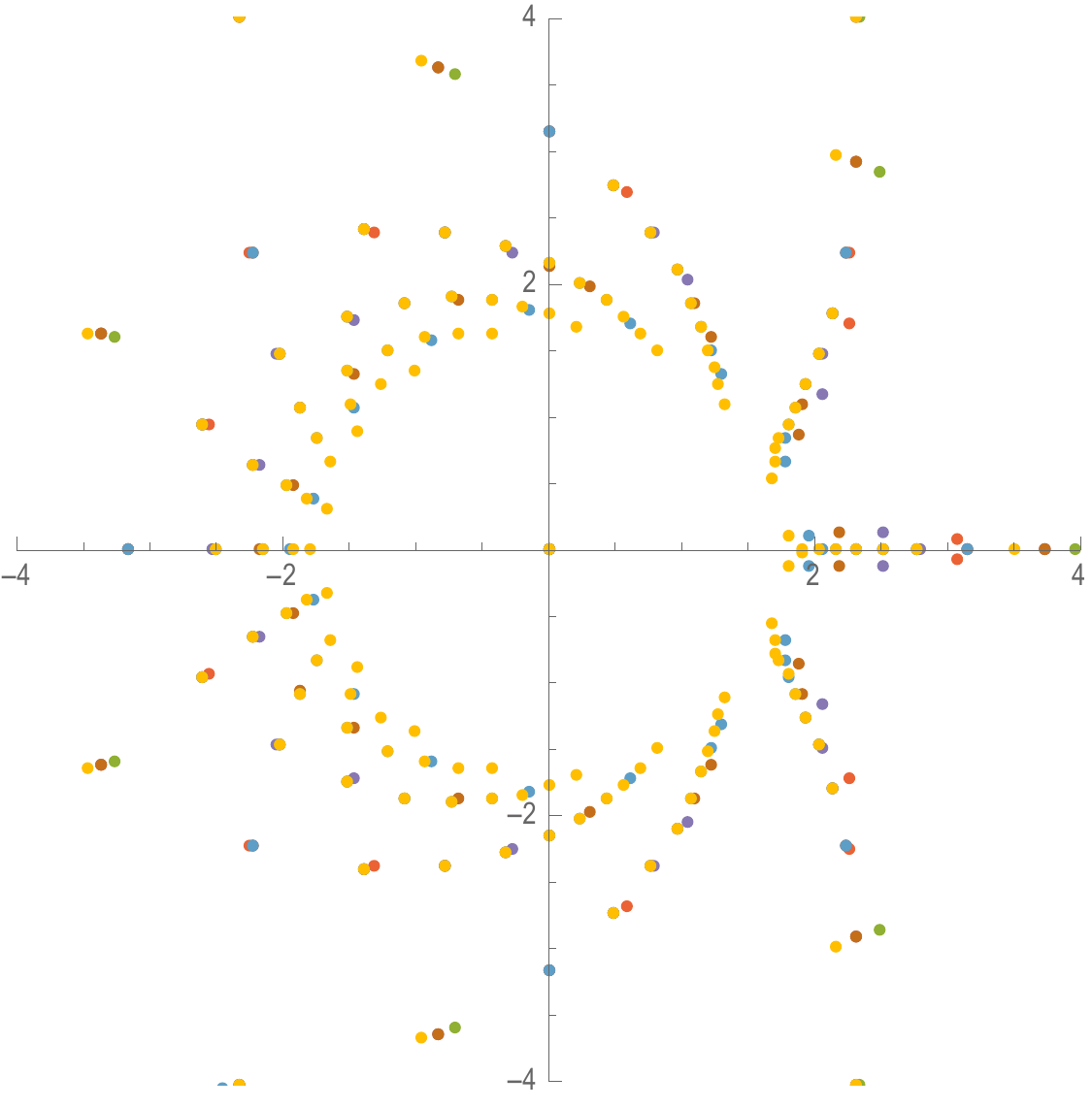} \ \ \ \ \
\includegraphics[scale=0.5]{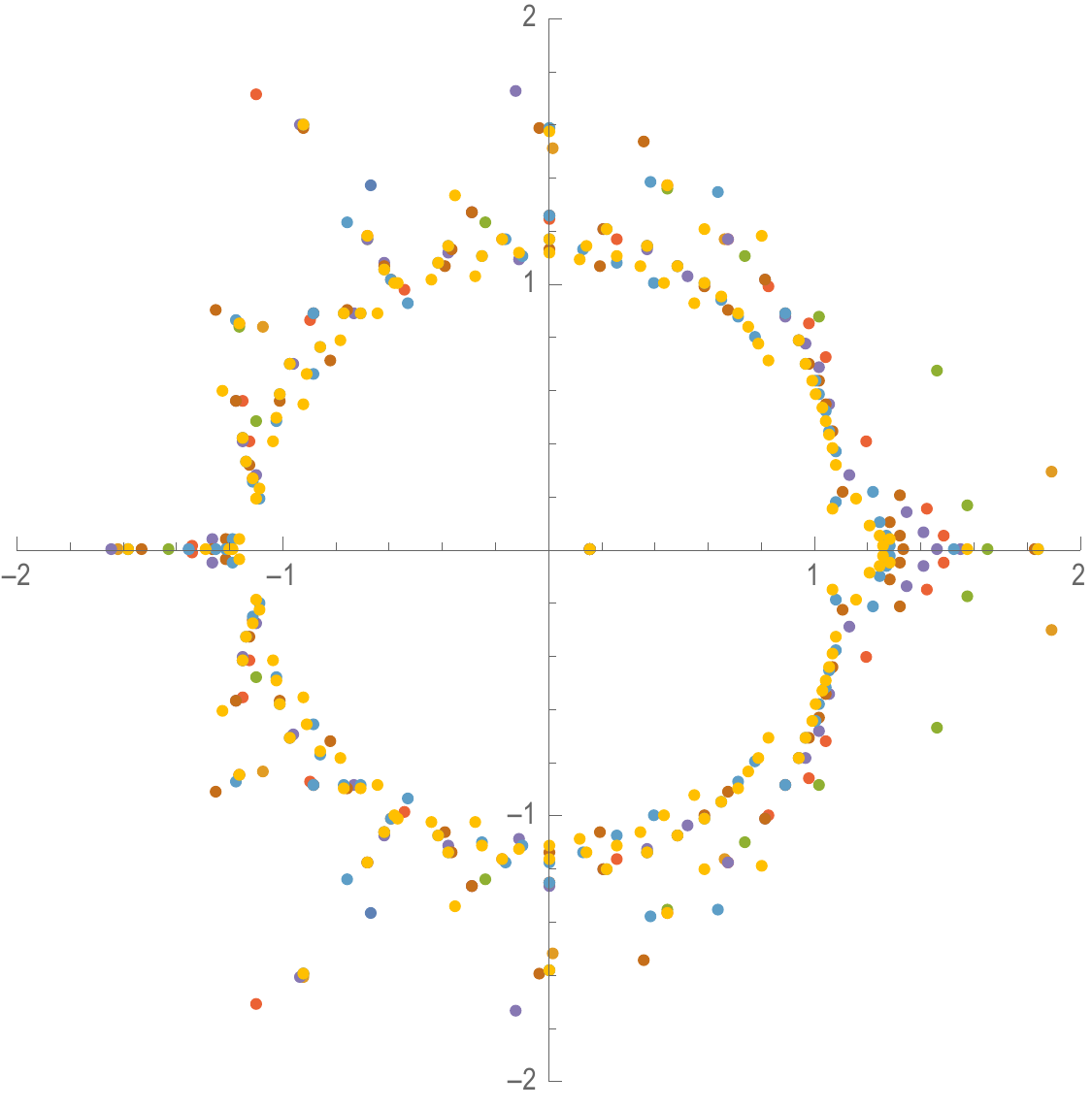}
}

\bigskip

$$
A = 0.9 \ \ \ \ \ \ \  \ \ \ \ \ \ \ \ \ \ \ \ \ \ \ \ \ \ \ \ \  \ \ \ \ \ \ \ \ \ \ \ \ \ \
A = 1-10^{-10}\ \ \ \ \ \ \  \ \ \ \ \ \ \ \ \ \ \ \ \ \ \ \ \ \ \ \ \  \ \ \ \ \ \ \ \ \ \ \ \ \ \
A=1
$$

\bigskip

\bigskip

\centerline{
\includegraphics[scale=0.5]{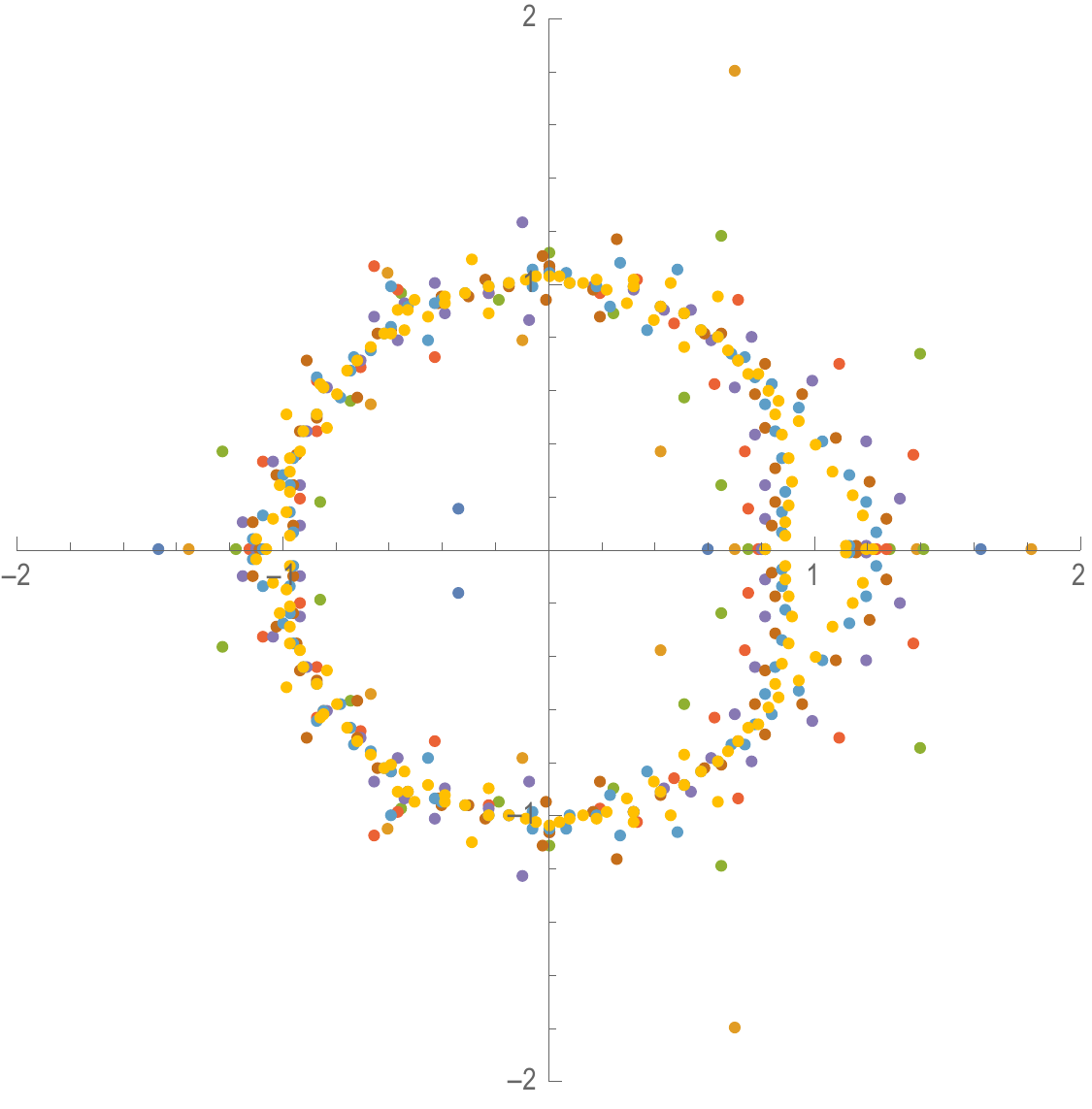} \ \ \ \
\includegraphics[scale=0.5]{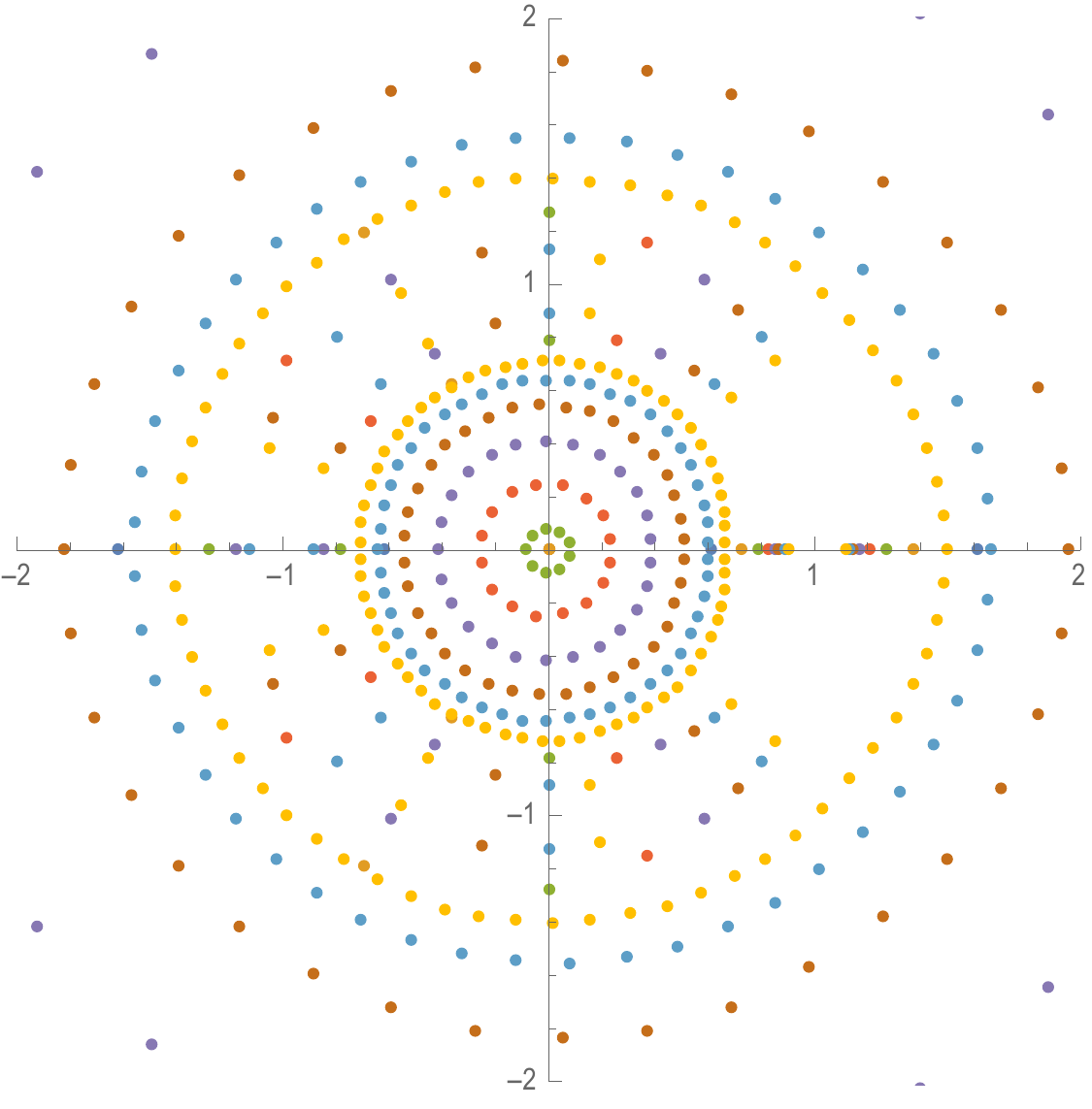} \ \ \ \
\includegraphics[scale=0.5]{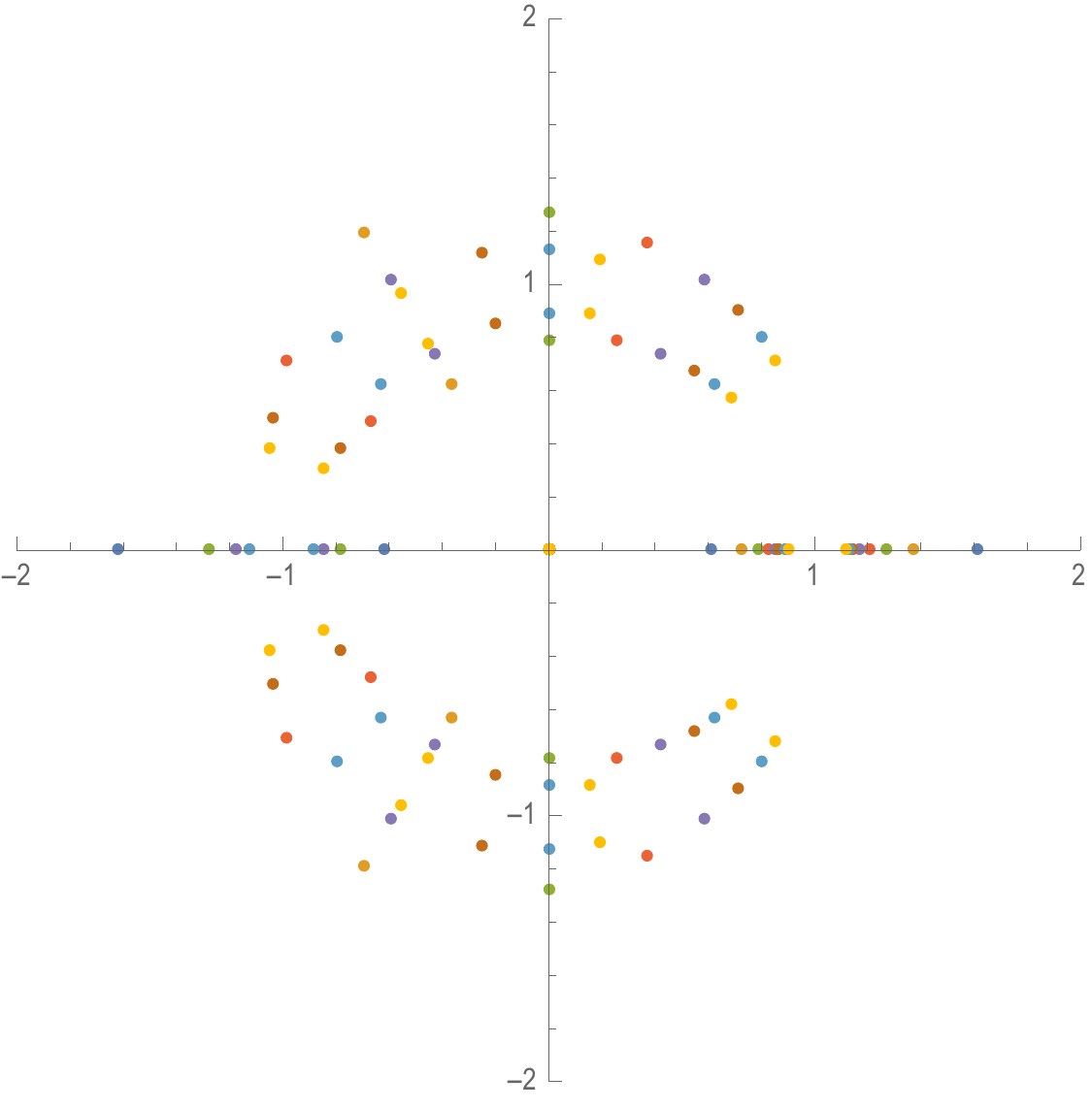}
}
\vspace{-10.5cm}
\be
\label{Fig15}
\ee
\vspace{8.6cm}

\bigskip

Immediately seen is the special role of the point $A^2=1$,
where the picture is drastically simplified.
This is not a surprise, because at this point
the roots of all symmetrically-colored $H_r$ are described by (\ref{Alex}).
Namely, they are all expressed through the roots $Q_\alpha$ of the
Alexander polynomial, whose degree is related to the defect $\delta^{\cal K}$ of the
differential expansion \cite{defect}:
\be
{\rm roots\ of\ } H_r(Q|A=\pm 1) =
\ \Big\{ \ Q_\alpha^{1/r} \cdot e^{2\pi i j/r}, \ \ \ \
\alpha = 1, \ldots, 2(\delta^{\cal K}+1), \ \ \ j=0,\ldots,r\ \Big\}
\ee
For a given $r$ the roots lie on concentric circles, shown in the picture is
a collection for various values of $r$ (different $r$ are shown in different colors)
-- thus the picture looks a little more complicated.

However,
a real complication appears when we leave the point $A^2=1$.
At $A^2\neq 1$ the power of $H_r$ in $Q$ is usually bigger than
$r\times$ that of the Alexander polynomial -- and in immediate vicinity
of $A=\pm 1$ the extra roots reappear from infinity and zero.
Thus the $r$-th component of the Julia set in the vicinity of $A=\pm 1$
has a peculiar form
of $2\delta^{\cal K}+4$ concentric circles round $Q=0$:
two of the $A$-dependent radii $\sim(A^2-1)^{\pm 1/\kappa_r}$
with $\kappa_r$ on each,
and $2\delta^{\cal K}+2$ circles of radii $Q_\alpha^{1/r}$ with
$2(\delta^{\cal K}+1)$ points on each.
Here
\be
\kappa_r = \ {\rm deg}_Q(H_r) - r(\delta^{\cal K}+1)
\ee
For some knots, including $3_1,5_1,5_2,7_1,7_2,7_3,7_4,7_5,\ldots$
all the Alexander roots $Q_\alpha$ belong to the unit circle --
then all the $A$-independent circles for all $\alpha$ and $r$ merge together
and get densely populated.

Despite universality of this pattern, it occurs only in the very close
vicinity of $A^2=1$ -- since $\kappa_r$ is typically large,
it breaks down at extremely small $A^2-1$, like $10^{-r}$ for $4_1$.
Indeed, the first zero of $discriminant_Q(H_r^{4_1})$, where some roots
from the extra two circles collides with those of original ones,
occurs already at
$A^2\approx 0.97$ and $A^2\approx 1.03$ for $r=3$.

\bigskip

When $A$ moves away from 1 concentric circles deform and collide with
"Alexander" circles (we plot zeros of $H_{10}^{4_1}$):

\bigskip

\bigskip

$$
A = 1 \ \ \ \ \ \ \ \ \ \ \ \ \ \ \ \ \ \ \ \ \  \ \ \ \ \ \ \ \ \ \ \ \ \ \ \ \ \ \ \ \ \
A = 1-10^{-10} \ \ \ \ \ \ \ \ \ \ \ \ \ \ \ \ \ \ \ \ \ \  \ \ \ \ \ \ \ \ \ \ \ \ \ \ \ \ \ \ \ \ \
A = 1-10^{-8}
$$

\bigskip

\centerline{
\includegraphics[scale=0.5]{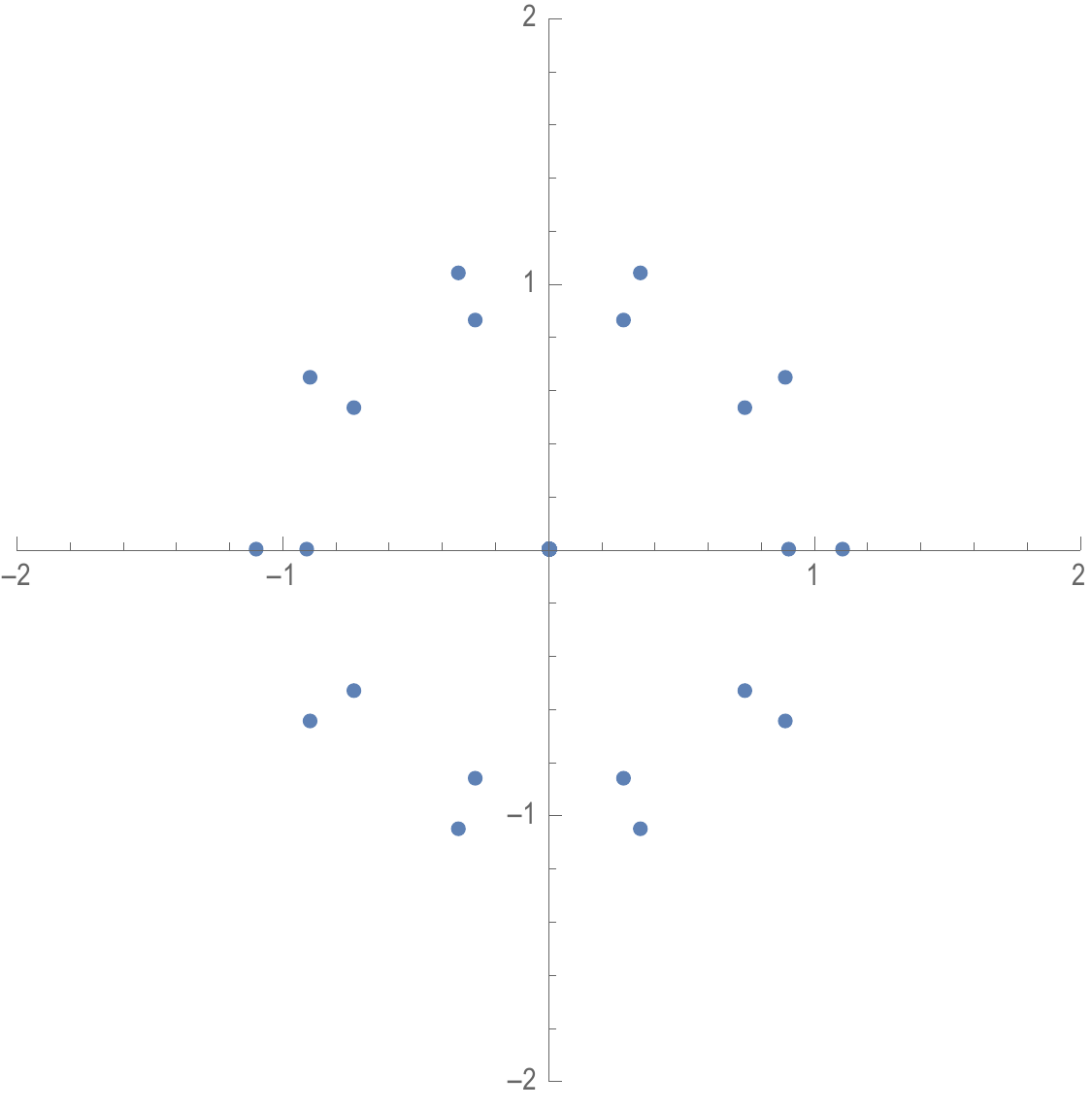} \  \  \
\includegraphics[scale=0.5]{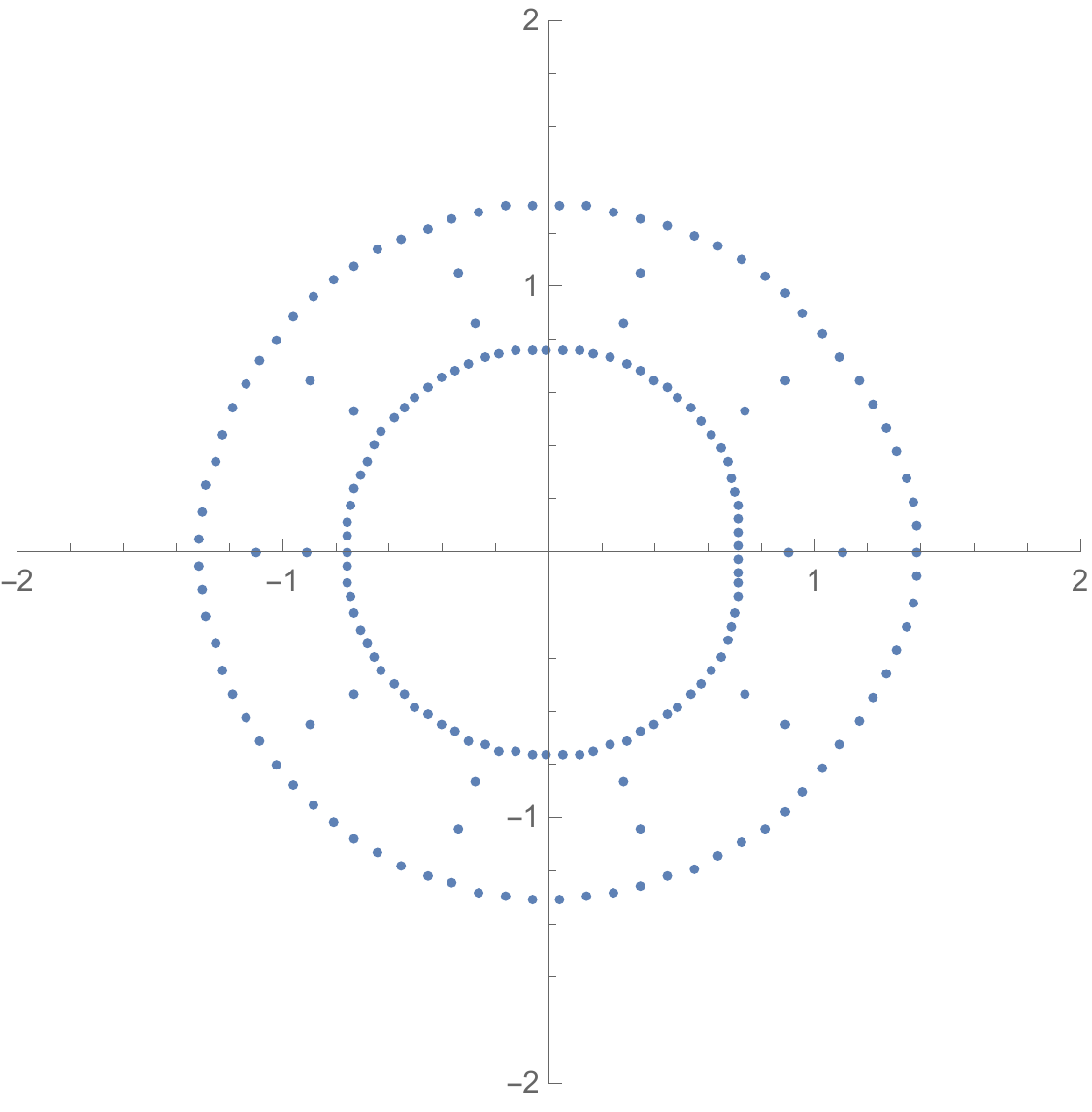} \  \  \
\includegraphics[scale=0.5]{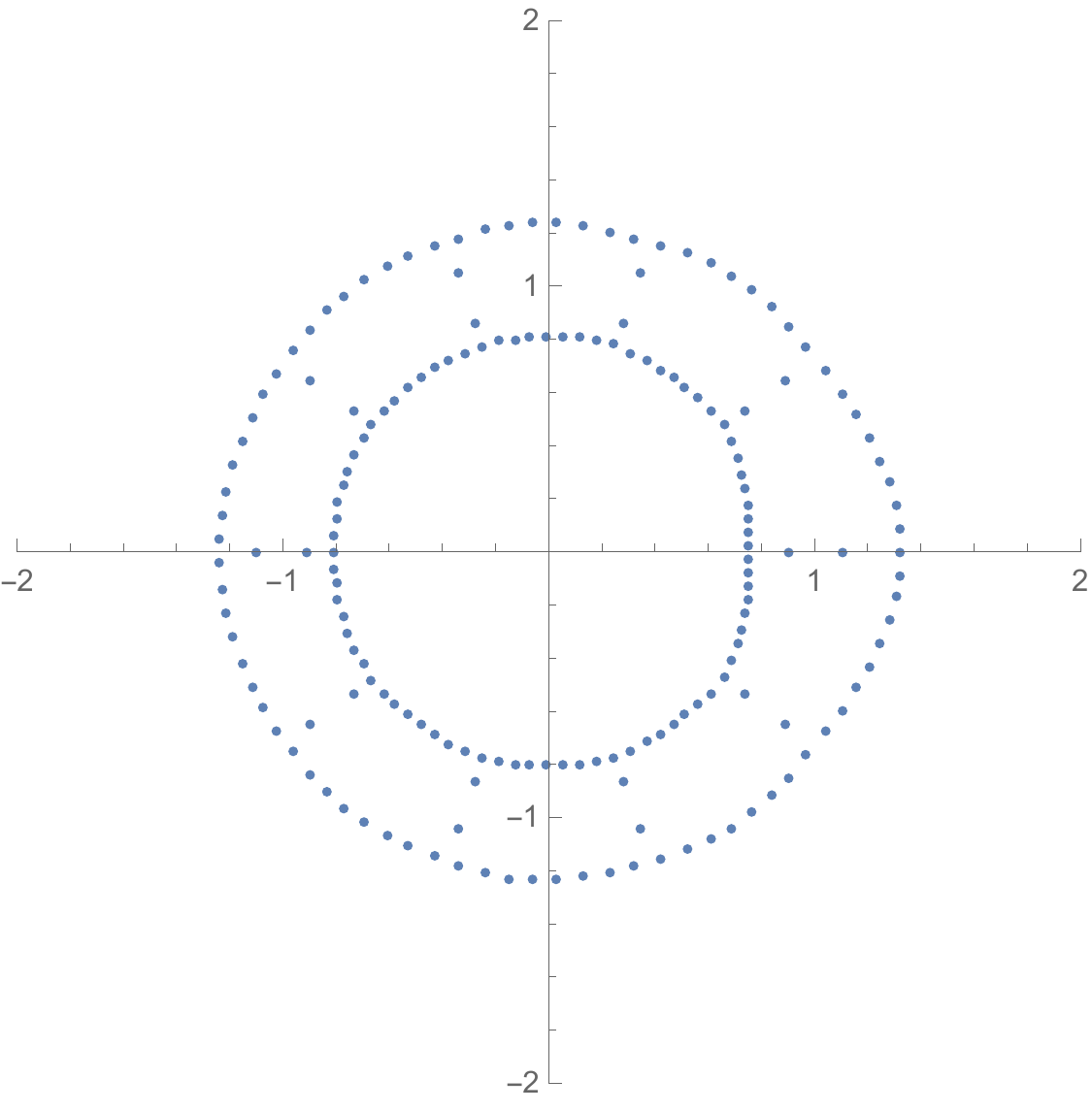}
}

\bigskip

\bigskip

$$
A = 1-10^{-6} \ \ \ \ \ \ \ \ \ \ \ \ \ \ \ \ \ \  \ \ \ \ \ \ \ \ \ \ \ \ \ \ \ \ \ \ \ \ \
A = 1-10^{-4} \ \ \ \ \ \ \ \ \ \ \ \ \ \ \ \ \ \  \ \ \ \ \ \ \ \ \ \ \ \ \ \ \ \ \ \ \ \ \
A = 1-10^{-1}
$$

\bigskip

\centerline{
\includegraphics[scale=0.5]{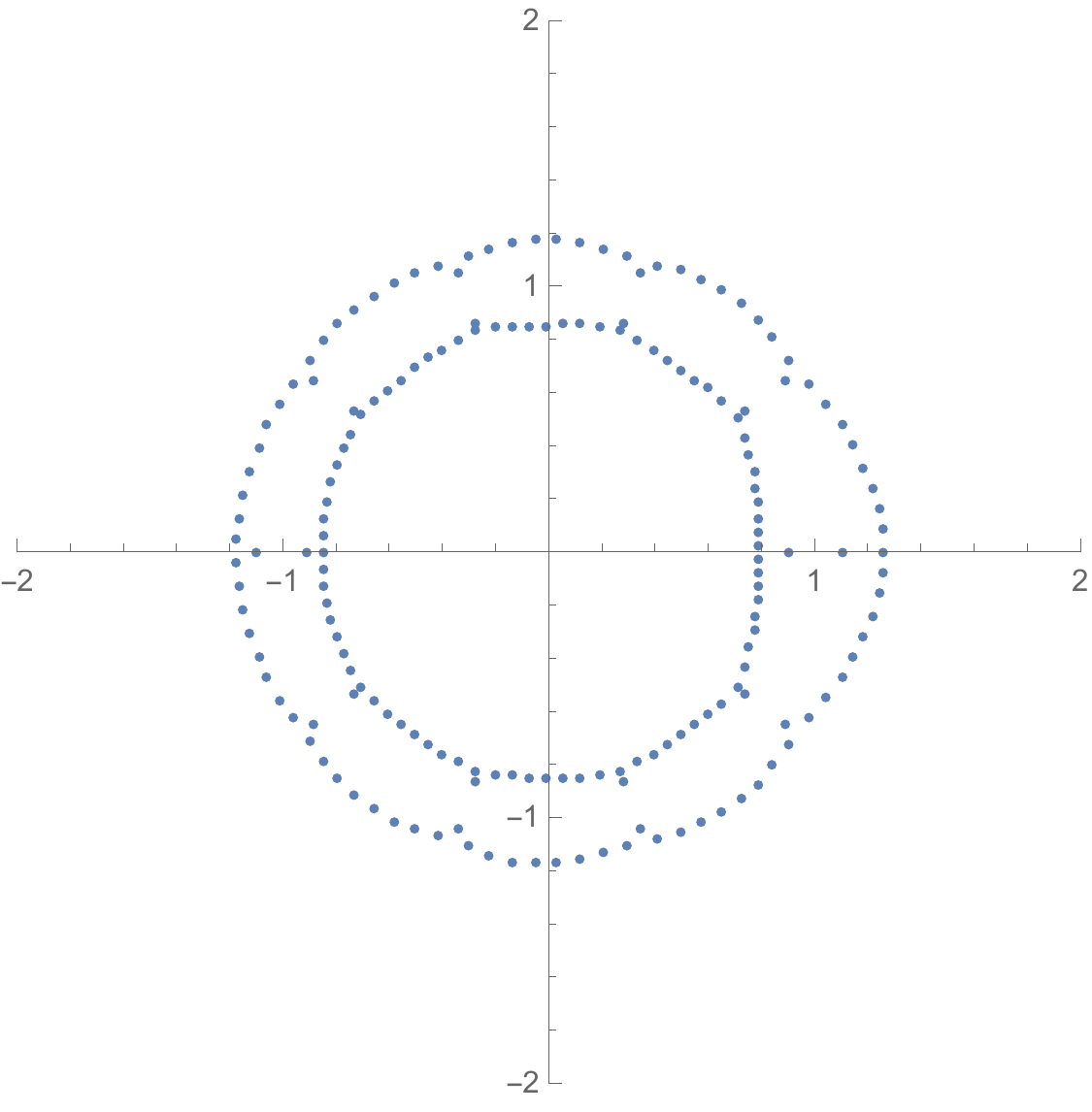}\  \  \
\includegraphics[scale=0.5]{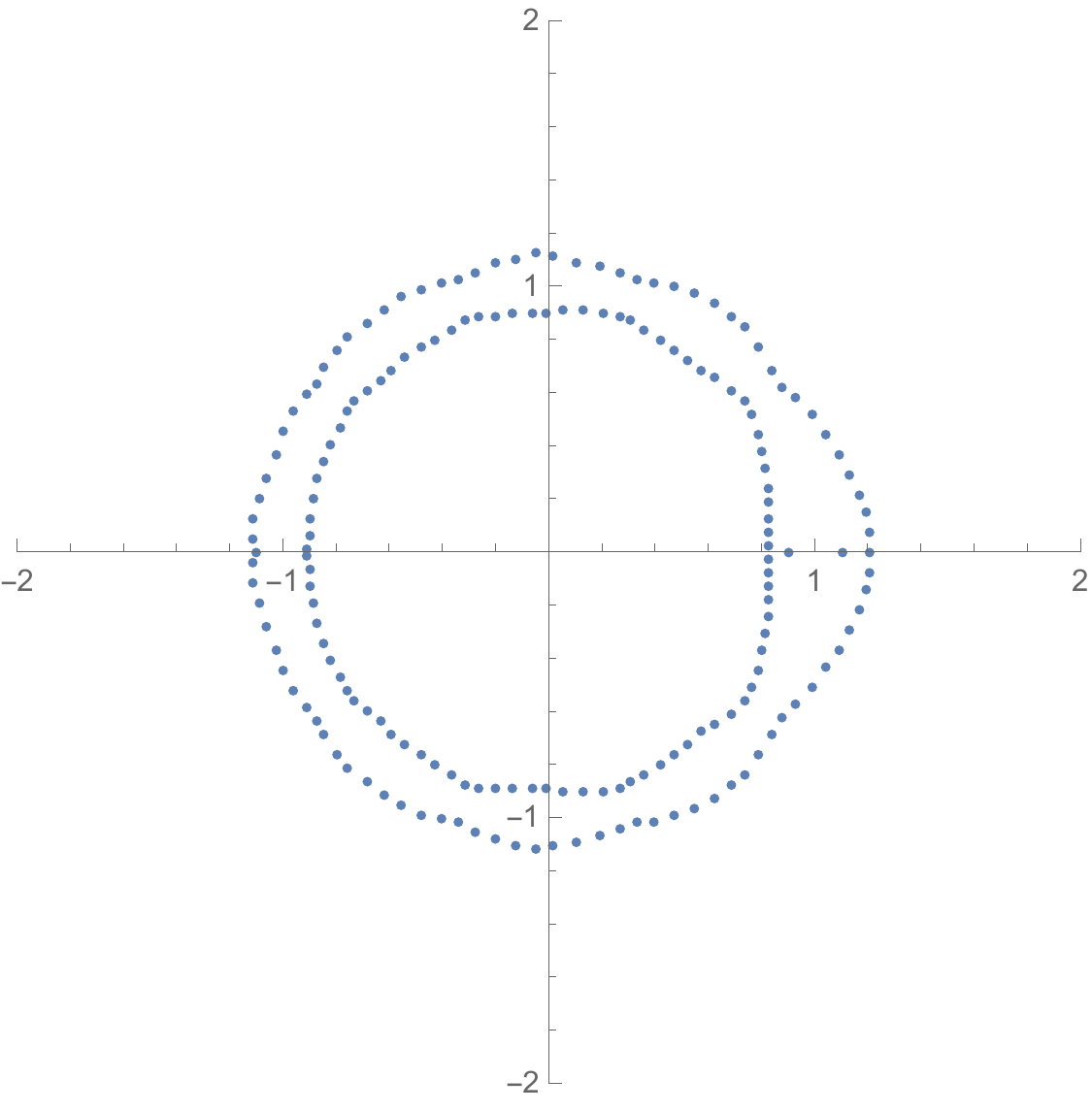} \  \  \
\includegraphics[scale=0.5]{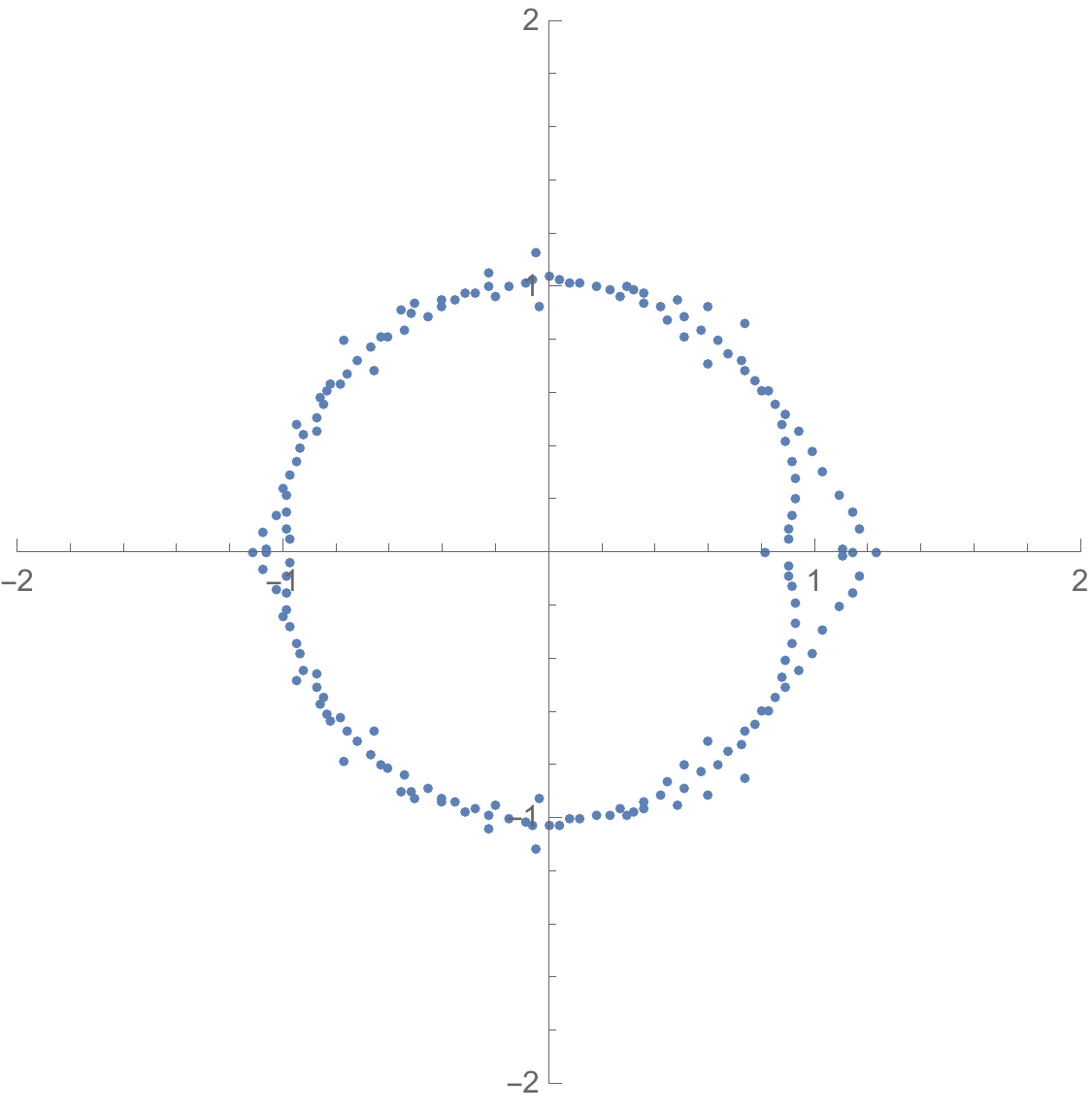}
}
\vspace{-9cm}
\be
\label{Fig16}
\ee
\vspace{7.1cm}

\bigskip

We can also consider perturbation in purely imaginary direction

\bigskip

$$
A = 1  \ \ \ \ \ \ \ \ \ \ \ \ \ \ \ \ \ \ \ \ \ \  \ \ \ \ \ \ \ \ \ \ \ \ \ \ \ \ \ \ \ \ \
A = 1-i \cdot 10^{-10} \ \ \ \ \ \ \ \ \ \ \ \ \ \ \ \ \ \ \ \ \ \  \ \ \ \ \ \ \ \ \ \ \ \ \ \ \ \ \ \ \ \ \
A = 1-i \cdot 10^{-8}
$$

\bigskip

\centerline{
\includegraphics[scale=0.5]{A=1.pdf} \  \  \
\includegraphics[scale=0.5]{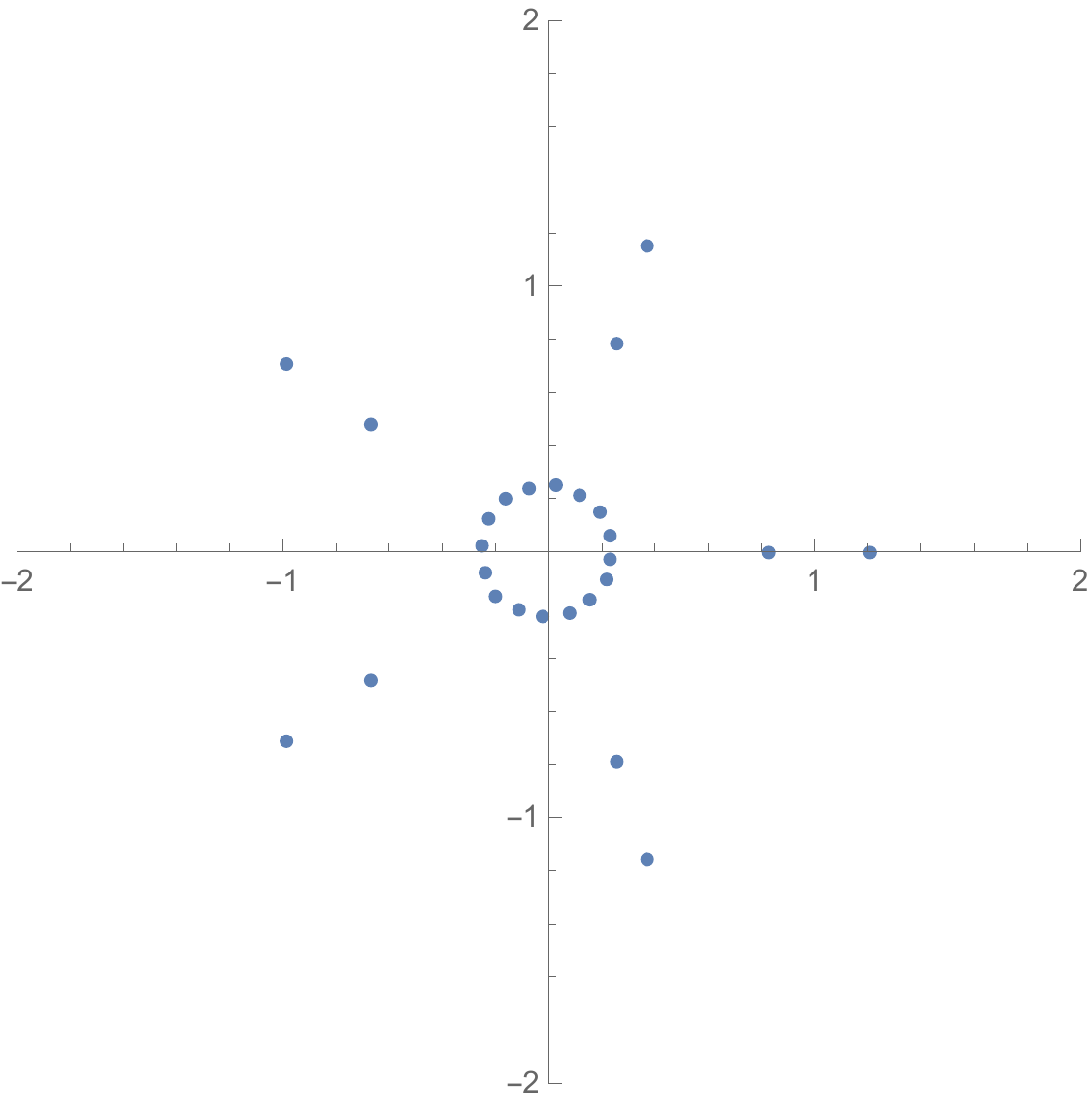} \  \  \
\includegraphics[scale=0.5]{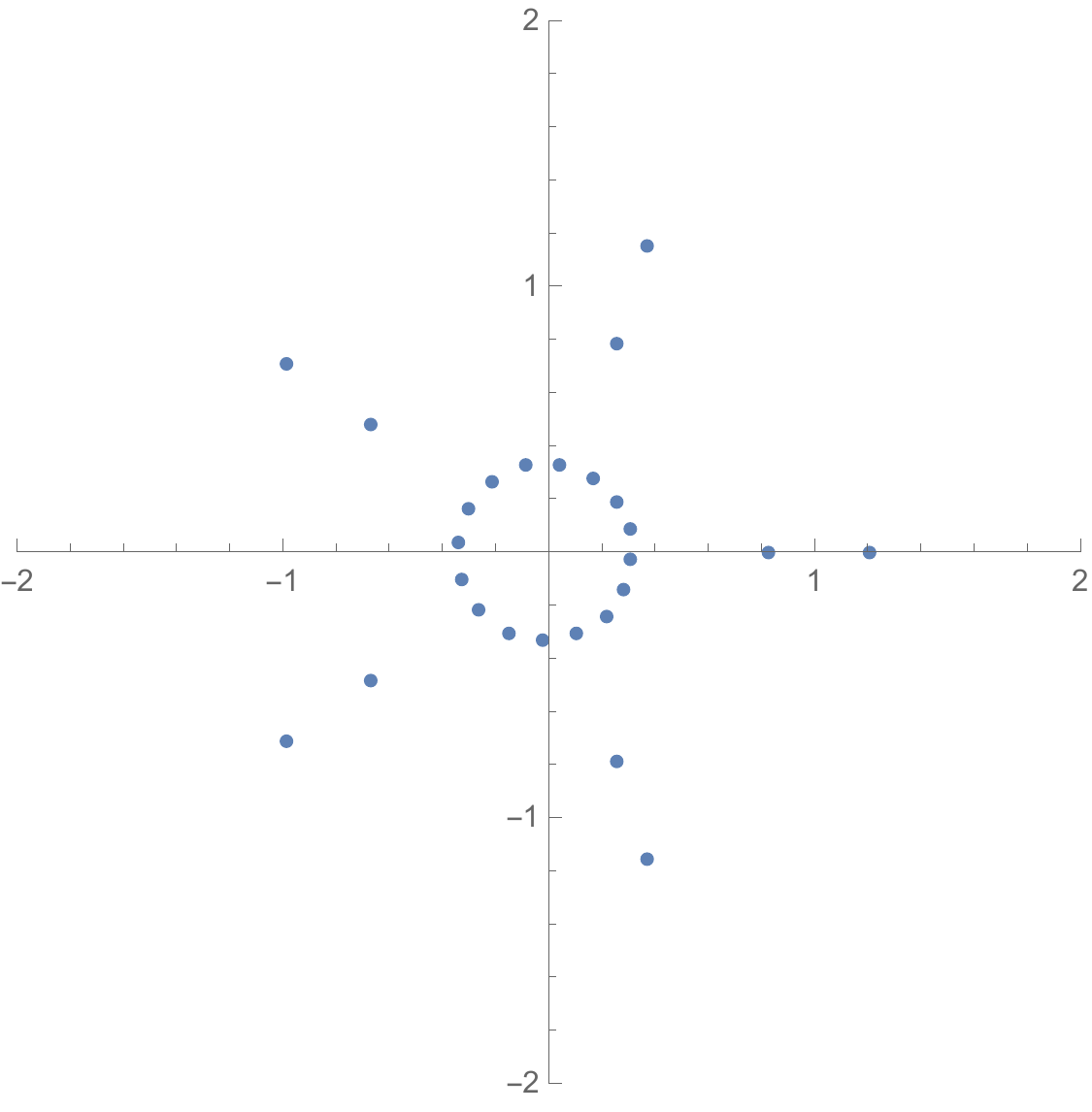}
}

\bigskip

$$
A = 1- i\cdot 10^{-6} \ \ \ \ \ \ \ \ \ \ \ \ \ \ \ \ \ \ \ \ \ \ \ \ \ \ \ \ \ \ \ \ \ \ \ \ \ \ \ \ \
A = 1-i\cdot 10^{-4} \ \ \ \ \ \ \ \ \ \ \ \ \ \ \ \ \ \ \ \  \ \ \ \ \ \ \ \ \ \ \ \ \ \ \ \ \ \ \ \ \
A = 1-i\cdot 10^{-1}\ :
$$

\bigskip

\centerline{
\includegraphics[scale=0.5]{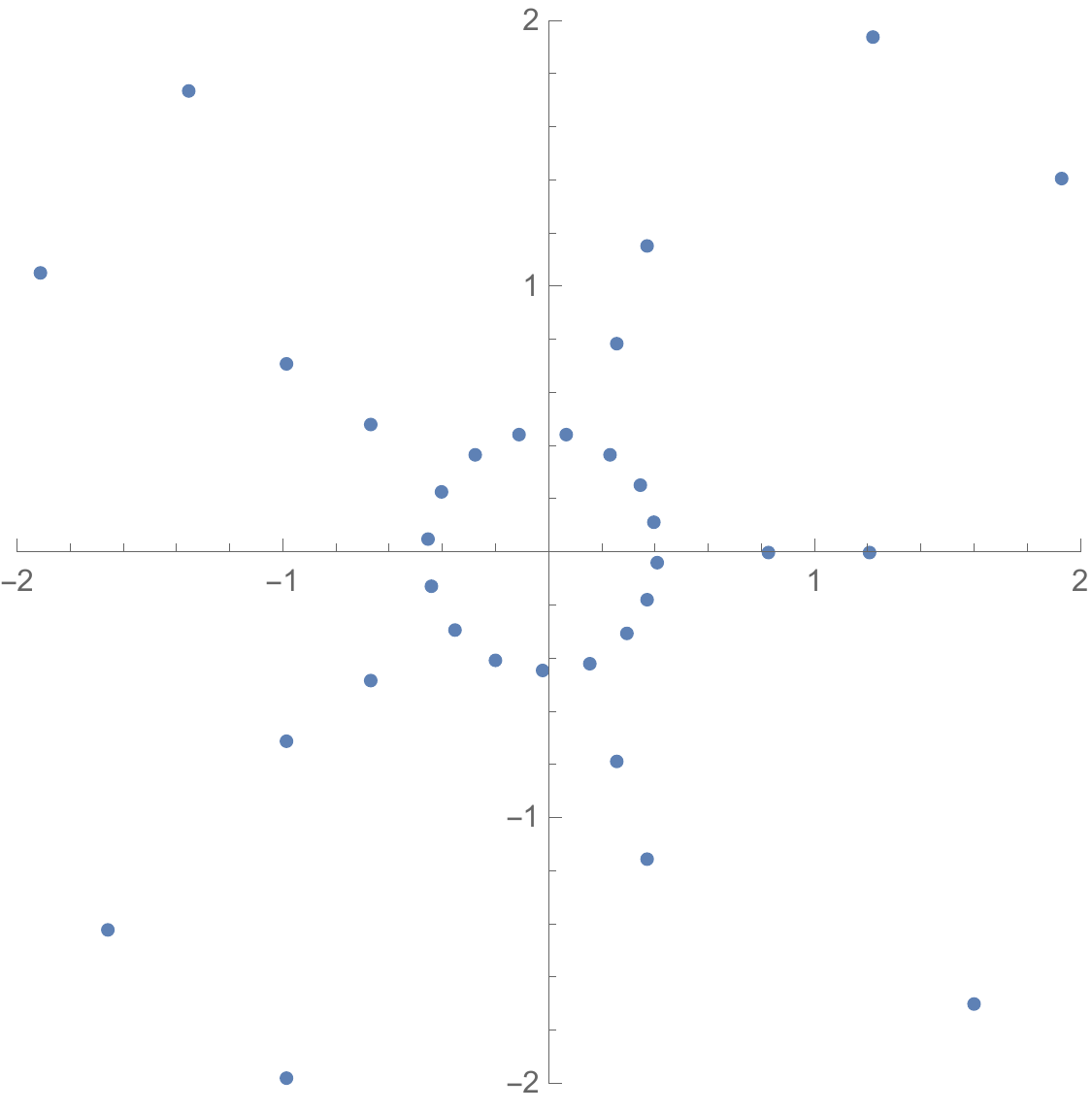} \  \  \
\includegraphics[scale=0.5]{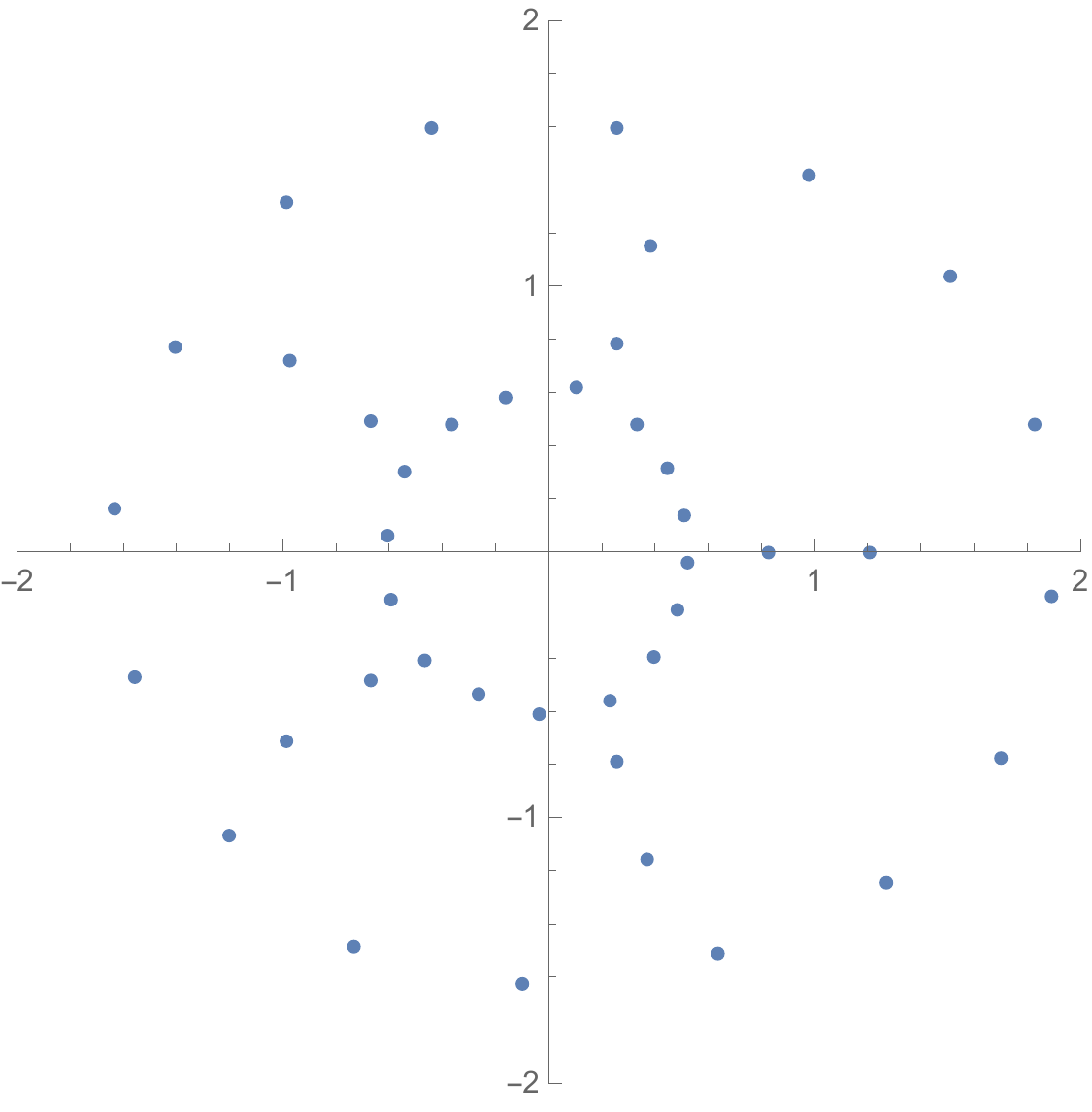} \  \  \
\includegraphics[scale=0.5]{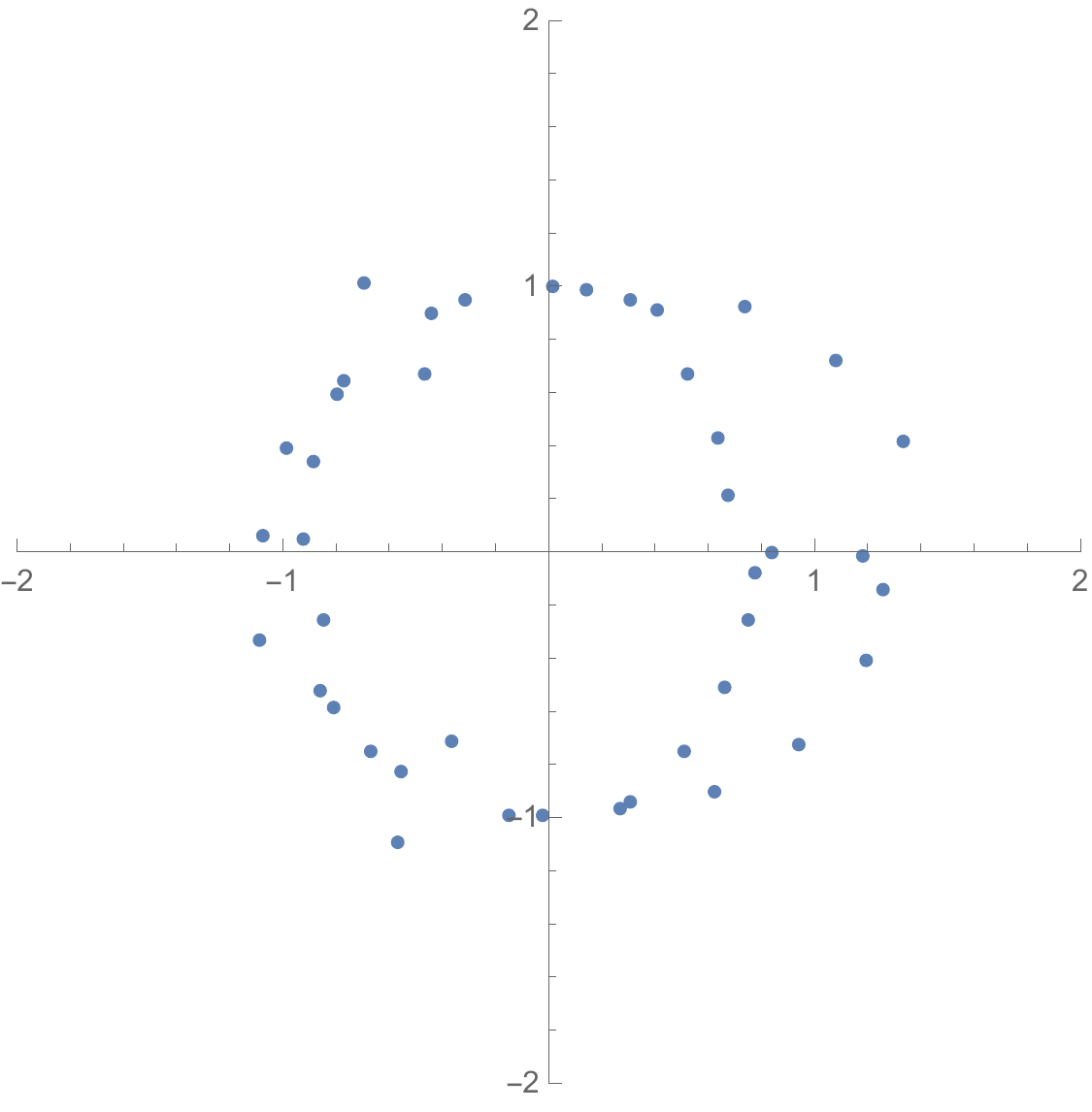}
}
\vspace{-9cm}
\be
\label{Fig16}
\ee
\vspace{7.1cm}

\bigskip

For the trefoil $3_1$ all the roots of Alexander polynomial are unimodular and some circles coincide:

\bigskip

$$
A=1 \ \ \ \ \ \ \ \ \ \ \ \ \ \ \ \ \ \ \ \ \ \  \ \ \ \ \ \ \ \ \ \ \ \ \ \ \ \ \ \ \ \ \
A = 1-10^{-10}
$$

\bigskip

\centerline{
\includegraphics[scale=0.5]{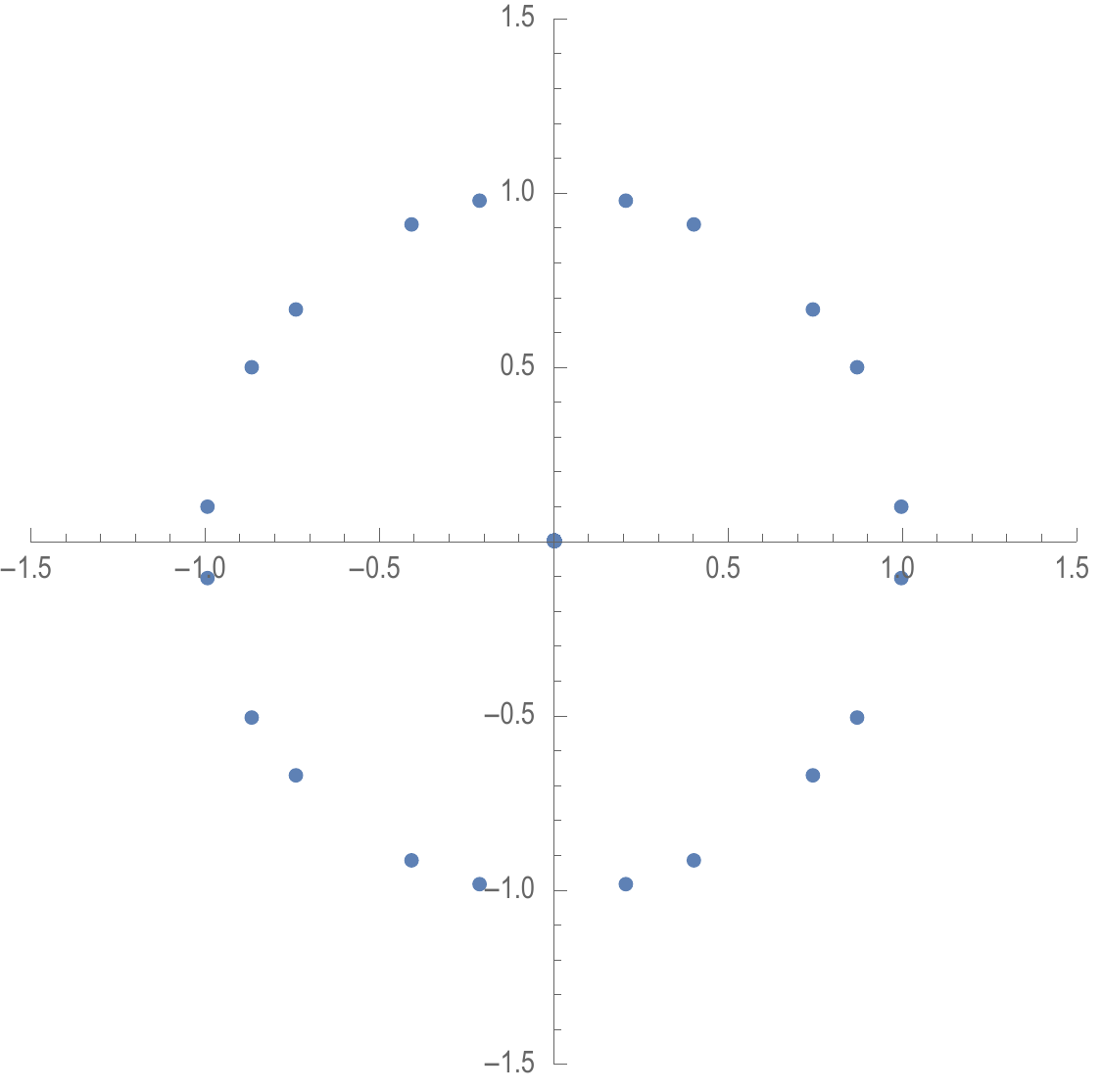} \  \  \
\includegraphics[scale=0.5]{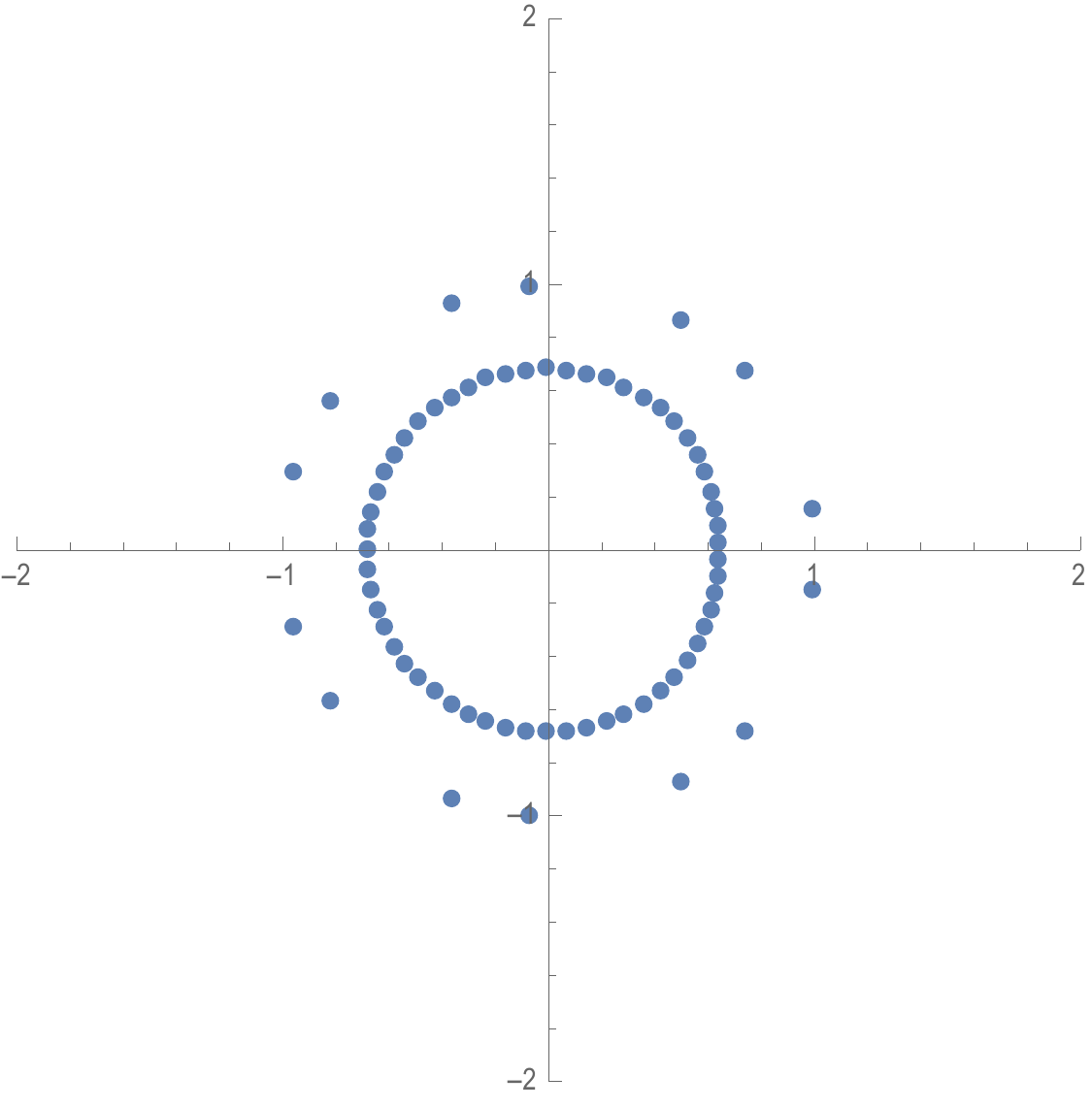}
}

\bigskip

$$
A = 1-10^{-4} \ \ \ \ \ \ \ \ \ \ \ \ \ \ \ \ \ \ \ \ \ \  \ \ \ \ \ \ \ \ \ \ \ \ \ \ \ \ \ \ \ \ \
A = 1-10^{-2} \ \ \ \ \ \ \ \ \ \ \ \ \ \ \ \ \ \ \ \ \ \  \ \ \ \ \ \ \ \ \ \ \ \ \ \ \ \ \ \ \ \ \
A = 1-10^{-1}
$$

\bigskip

\centerline{
\includegraphics[scale=0.5]{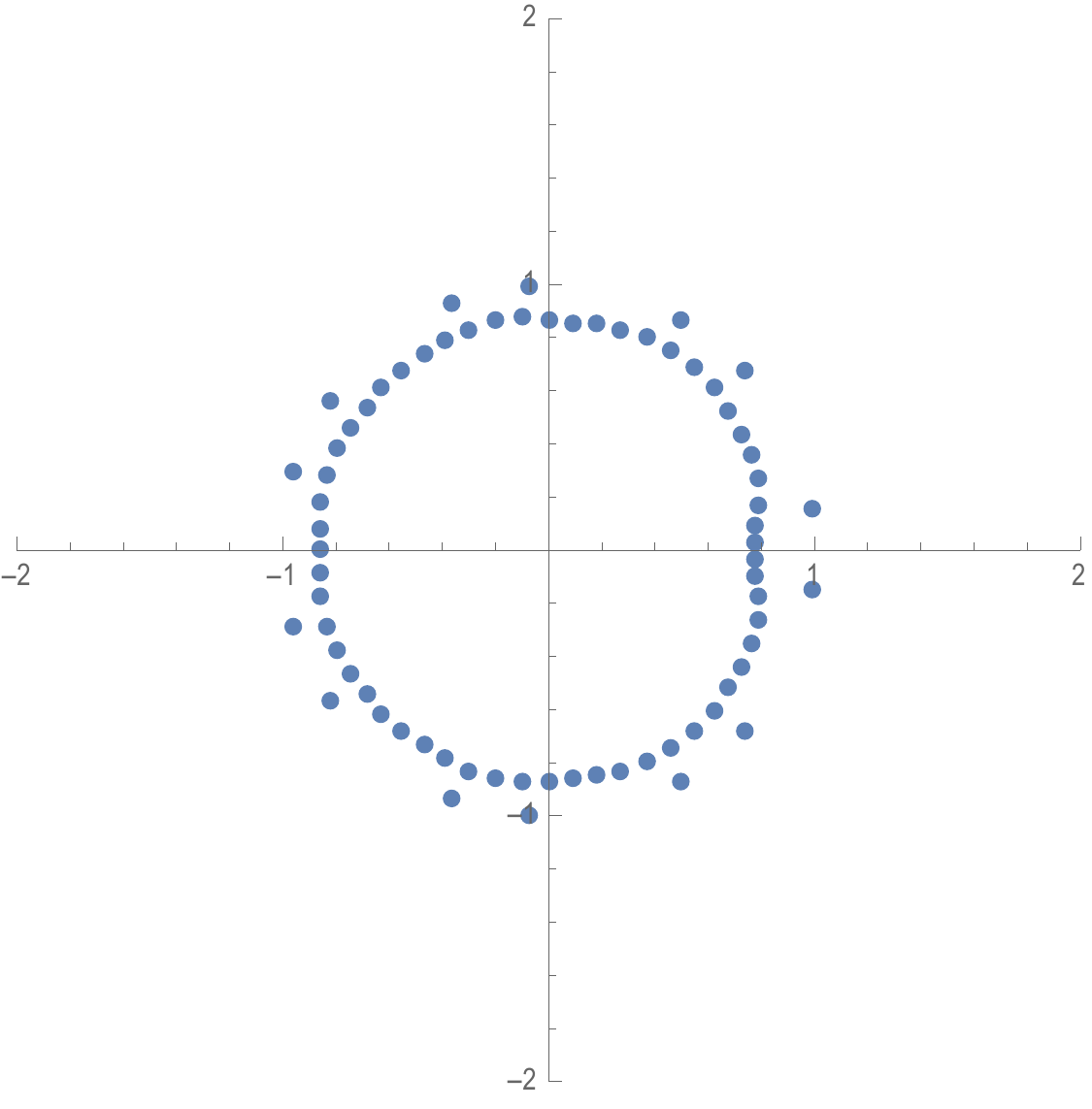} \  \  \
\includegraphics[scale=0.5]{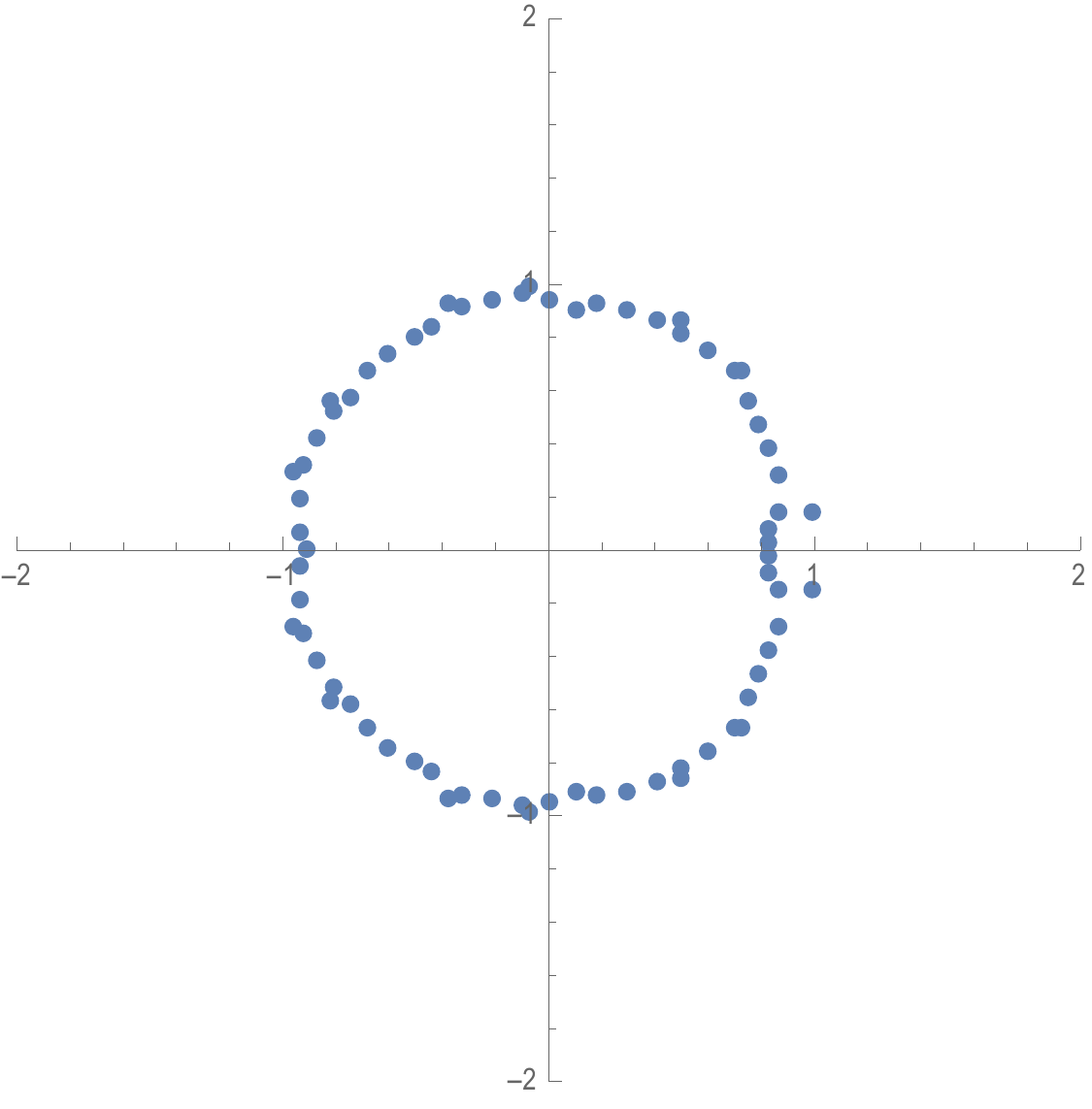} \  \  \
\includegraphics[scale=0.5]{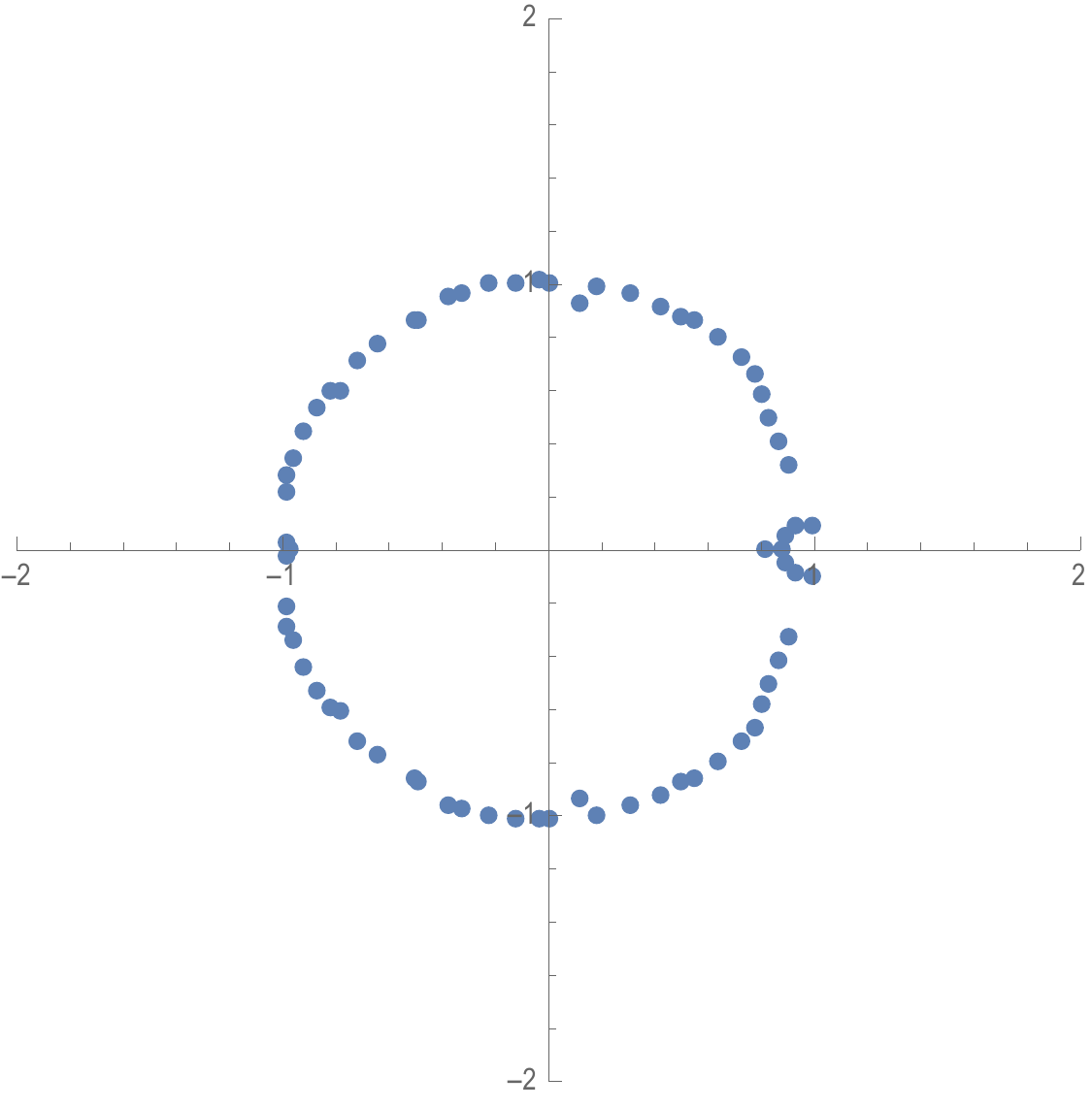} \  \  \
}

\bigskip

\noindent
and

\bigskip

$$
A = 1 \ \ \ \ \ \ \ \ \ \ \ \ \ \ \ \ \ \ \  \ \ \ \ \ \ \ \ \ \ \ \ \ \ \ \ \ \ \ \ \
A = 1-i\cdot 10^{-10}\ \ \ \ \ \ \ \ \ \ \ \ \ \ \ \ \ \ \ \  \ \ \ \ \ \ \ \ \ \ \ \ \ \ \ \ \ \ \ \ \
A = 1-i\cdot 10^{-4}
$$

\bigskip

\centerline{
\includegraphics[scale=0.5]{31A=1.pdf} \  \  \
\includegraphics[scale=0.5]{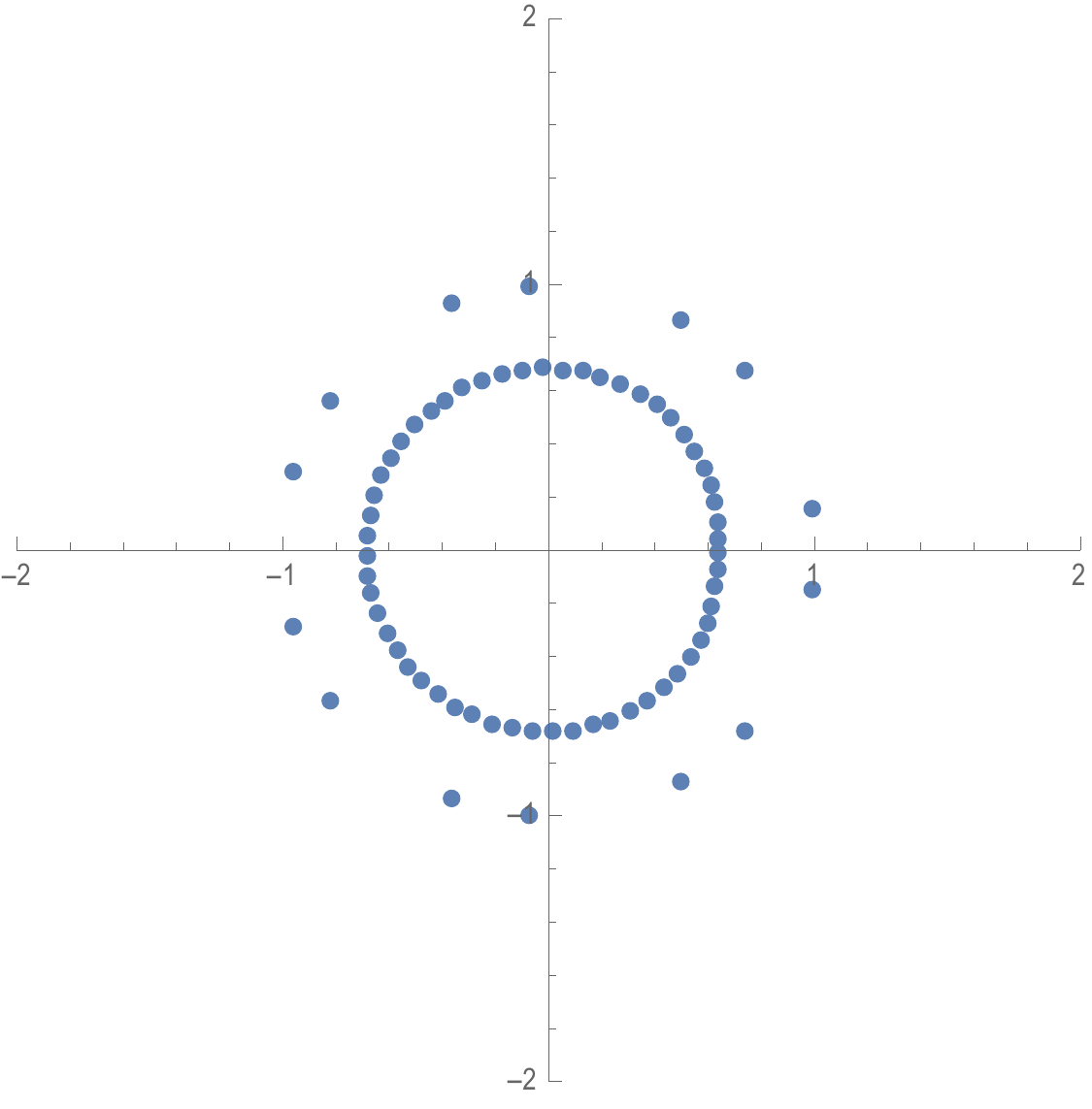} \  \  \
\includegraphics[scale=0.5]{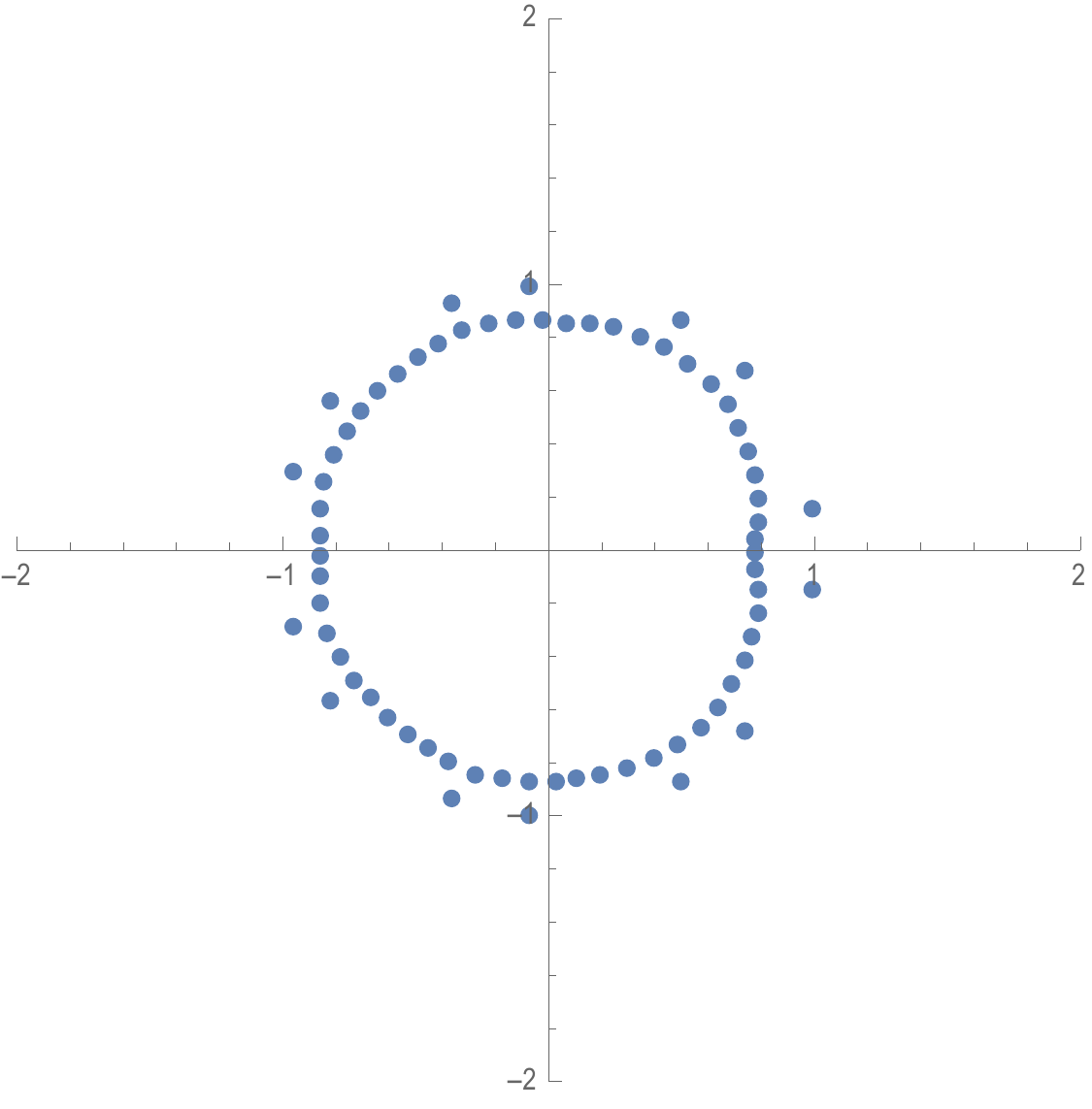}
}

\bigskip

$$
A = 1-i\cdot 10^{-2} \ \ \ \ \ \ \ \ \ \ \ \ \ \ \ \ \ \ \ \
A = 1-i\cdot 10^{-1} \ \ \ \ \ \ \ \ \ \ \ \ \ \ \ \ \ \ \ \
A = 1-i
$$

\bigskip

\centerline{
\includegraphics[scale=0.5]{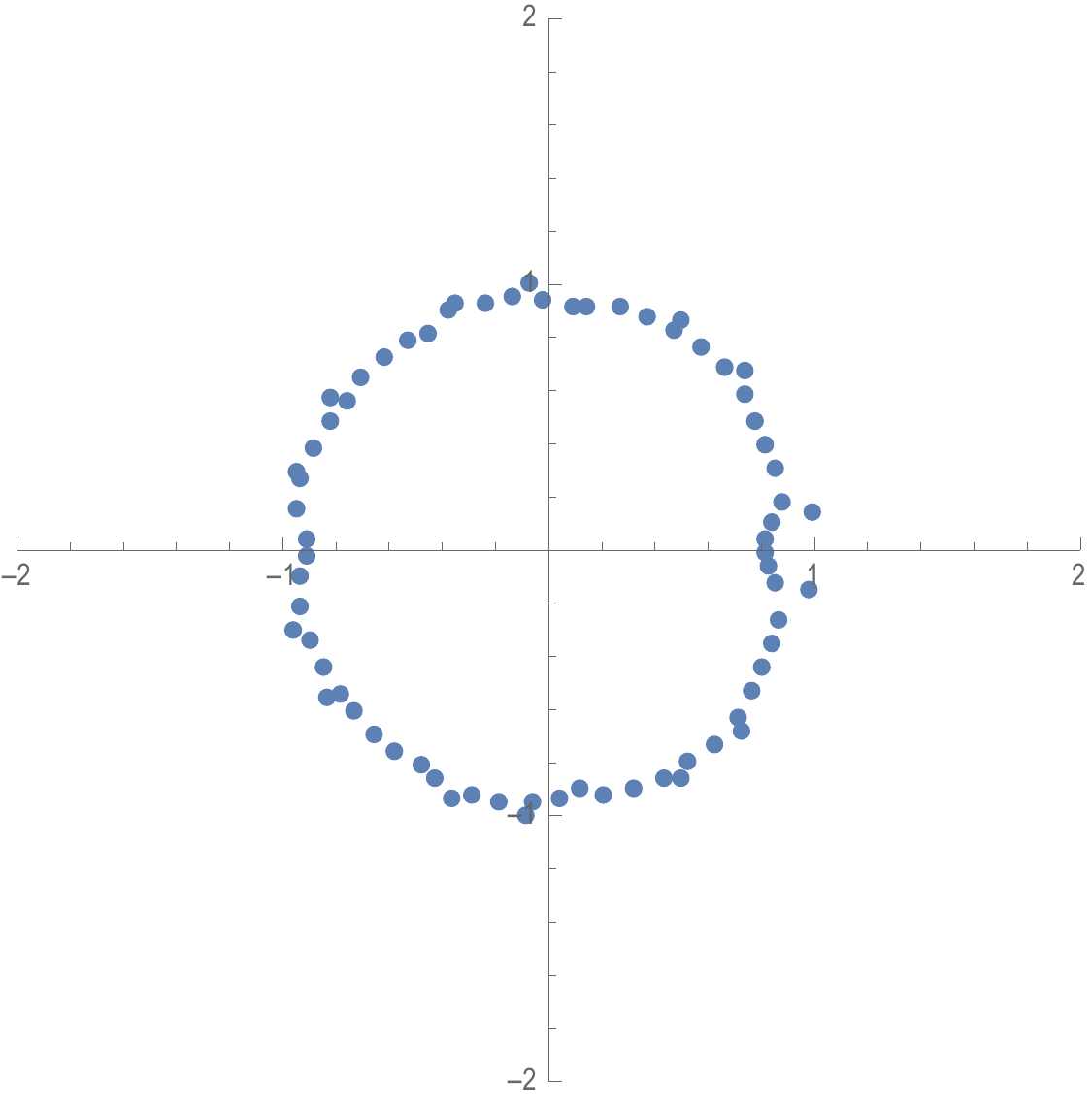} \  \  \
\includegraphics[scale=0.5]{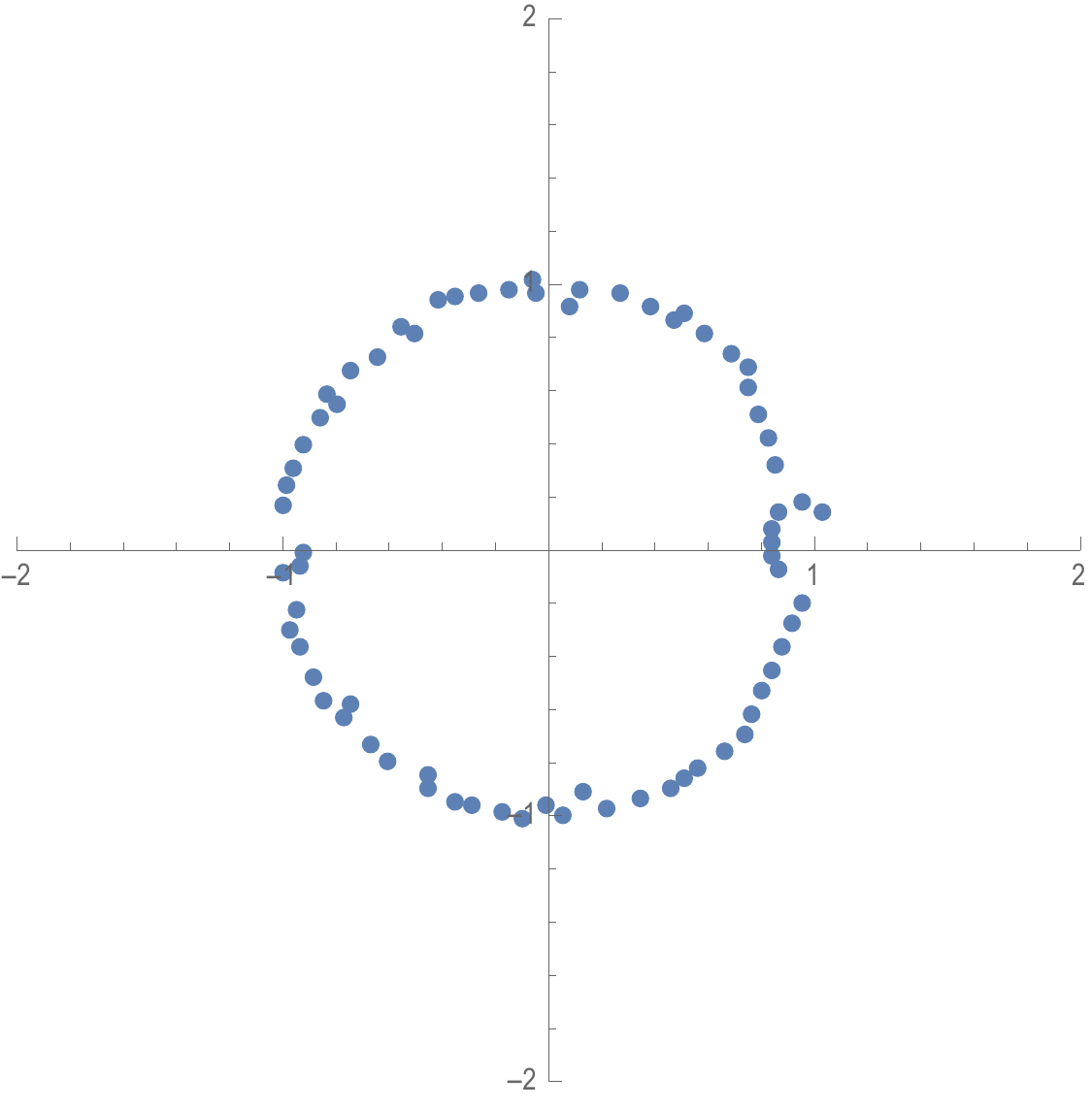} \  \  \
\includegraphics[scale=0.5]{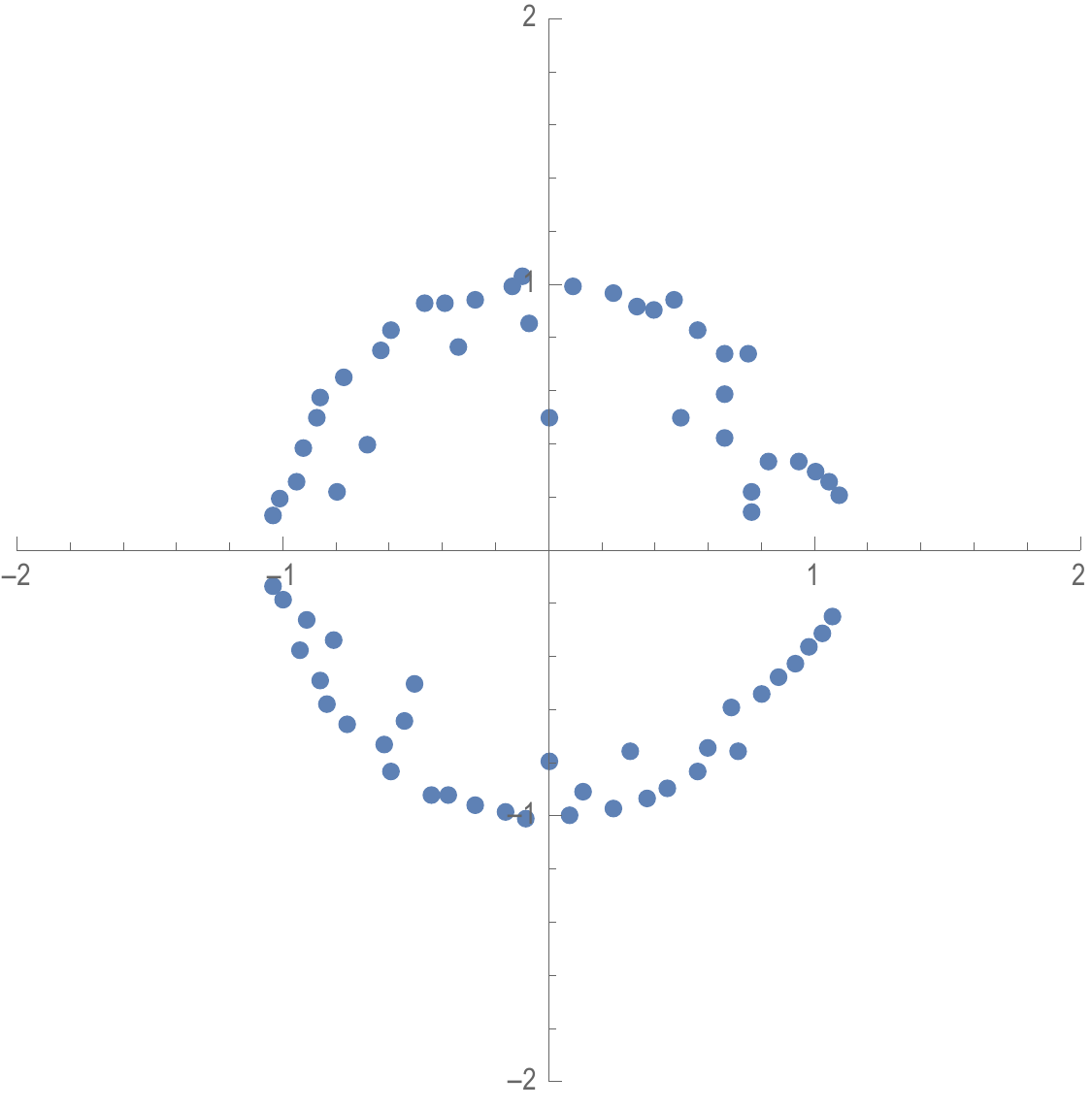}
}
\vspace{-16.5cm}
\be
\label{Fig17}
\ee
\vspace{14.6cm}

It turns out that if $\{q^r\} = 1$ then
\be
\left( \frac{d}{dA} (H_r) \right)_{A=1} = 0.
\ee
This allows us to define
\be
H_r = Al(q^r) + \{A\}\{q^r\} P(q) + o(A-1).
\ee
The behavior of concentric circles depends much on the degrees
of $Al(q^r)$ and $P(q)$.
For $3_1$-knot it turns out that each
term of expansion with respect to $A-1$ contains the same
maximal positive power of $q$, but different negative powers.
This is the reason to the fact that there exists a circle collapsing to the origin,
but there is not a circle shrinking to infinity.

\bigskip

We also clearly see a cavity near the point $q=1$. For the $4_1$ the picture looks like
(we show also the unimodular circle for convenience):

\bigskip

\centerline{
\includegraphics[scale=0.5]{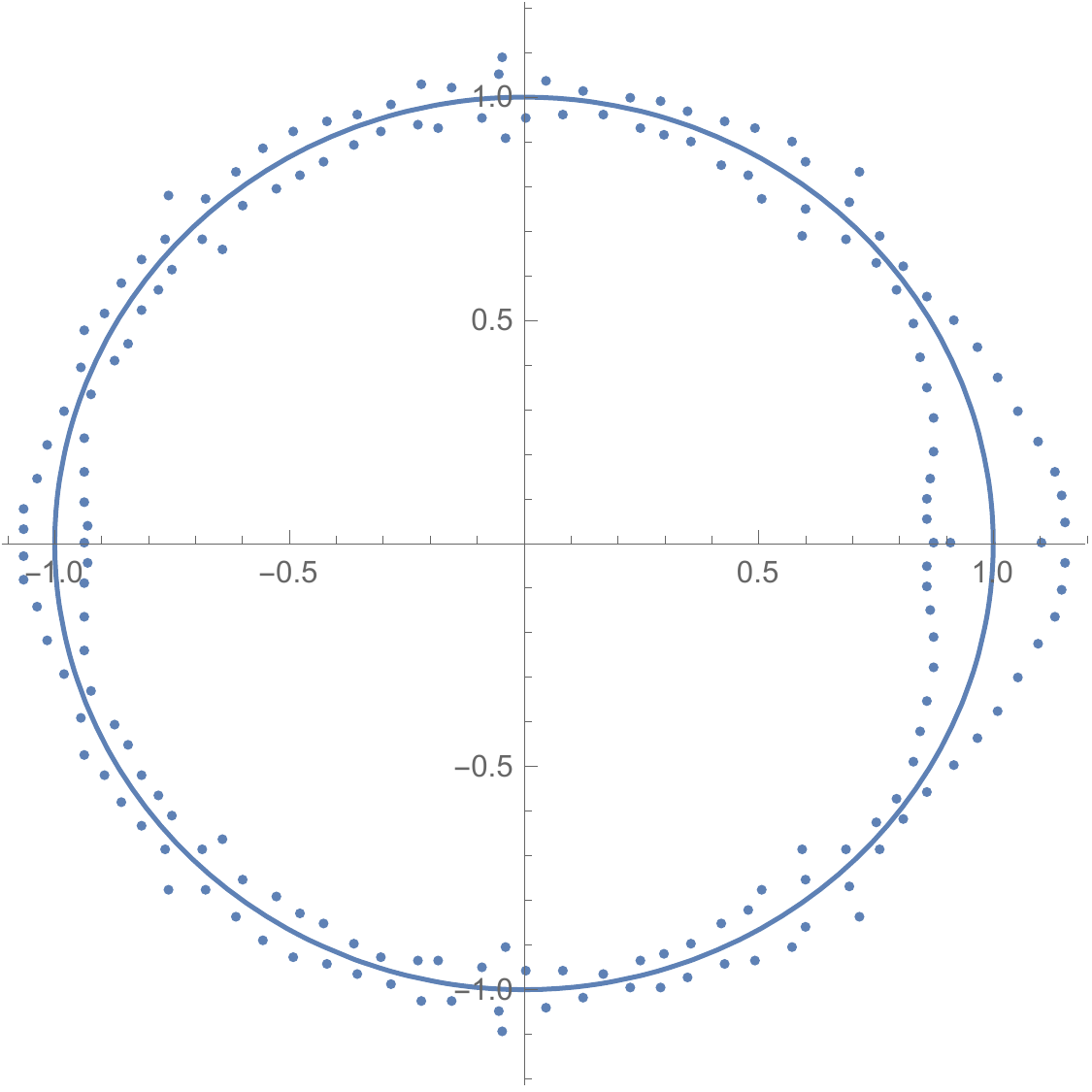}
}
\vspace{-4cm}
\be
\label{Fig18}
\ee
\vspace{2.1cm}

However, when the degrees coincide (say, for $3_1$), the picture looks differently:

\bigskip

\centerline{
\includegraphics[scale=0.5]{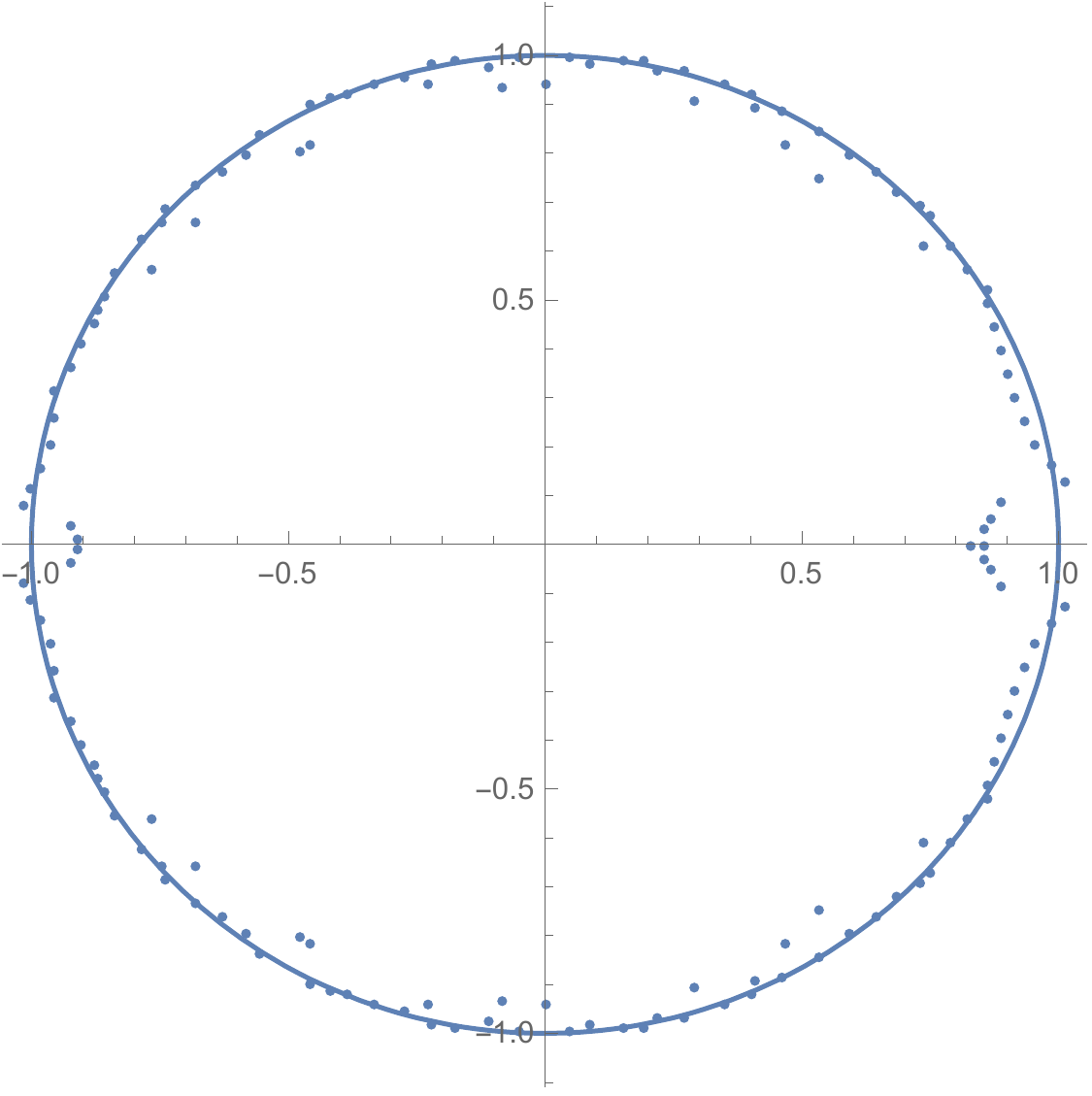}
}
\vspace{-4cm}
\be
\label{Fig19}
\ee
\vspace{2.1cm}

\section{Conclusion}

In this paper we provided some evidence that colored HOMFLY polynomials
can exhibit the what we called Mandelbrot property: zeroes of
$resultant_{q^2} (H_m,H_n)$ form structures of codimension one
in the complex-$A$ plane.
Accuracy of this statement is questionable: emerging curves are at best
not as smooth as they were for Mandelbrot sets -- nothing to say that we
have not any analytical description of those, similar to (\ref{Mandanalyt}).
Correlation between zeroes at different $n$ is clearly present, but it
still a question, if it is sufficient to {\it exactly} provide the Mandelbrot property.
Moreover, the quality of "one-dimensionality" seems related to other
properties of knots -- for example, it looks better for knots with defect zero.

In fact, the choice of HOMFLY $H_n$ instead of, say, the differences $H_n-1$,
for the substitute of the shifted iterated maps,
or of $A$ for the role of modulus  are rather arbitrary, and these choices
affect the pictures -- though not as much as one could expect
(what is, however, in full
accordance with the viewpoint of \cite{DMand}, thus only confirming the possible
relation to Mandelbrot theory).

Further research is needed to understand what really happens.
Straightforward computer experiments are not easy -- calculation of complicated
colored HOMFLY is still a challenge, and finding zeroes of resultants of
complicated polynomials is also a difficult task for today's computers.
What is needed is better theoretical understanding of HOMFLY polynomials
and especially of their factorization properties and zeroes -- perhaps,
in the spirit of our simple observations in \cite{Hunicircle}.

\section*{Acknowledgements}

 This work was performed at the Institute for Information Transmission Problems with the financial
support of the Russian Science Foundation (Grant No.14-50-00150).

\end{document}